\documentclass[10pt,a4paper,twocolumn]{article}

\usepackage{times}

\usepackage{amsmath}
\usepackage{amsfonts}
\usepackage{amssymb}
\usepackage{graphicx}
\usepackage{color}
\usepackage{url}
\usepackage{animate}
\usepackage[ruled]{algorithm2e}
\usepackage{setspace}
\usepackage{stfloats}
\usepackage{titlesec}

\makeatletter
\def\UrlAlphabet{%
      \do\a\do\b\do\c\do\d\do\e\do\f\do\g\do\h\do\i\do\j%
      \do\k\do\l\do\m\do\n\do\o\do\p\do\q\do\r\do\s\do\t%
      \do\u\do\v\do\w\do\x\do\y\do\z\do\A\do\B\do\C\do\D%
      \do\E\do\F\do\G\do\H\do\I\do\J\do\K\do\L\do\M\do\N%
      \do\O\do\P\do\Q\do\R\do\S\do\T\do\U\do\V\do\W\do\X%
      \do\Y\do\Z}
\def\UrlDigits{\do\1\do\2\do\3\do\4\do\5\do\6\do\7\do\8\do\9\do\0}
\g@addto@macro{\UrlBreaks}{\UrlOrds}
\g@addto@macro{\UrlBreaks}{\UrlAlphabet}
\g@addto@macro{\UrlBreaks}{\UrlDigits}

\vspace{0cm} 
\titlespacing*{\section} {0pt}{6pt}{3pt}
\titlespacing*{\subsection} {0pt}{5pt}{3pt}

\newenvironment{sciabstract}{%
\begin{quote} \bf}
{\end{quote}}

\title{Public discourse and social network echo chambers driven by socio-cognitive biases}

\author
{Xin Wang,${}^{1,2}$ Antonio D. Sirianni,${}^{3\dagger}$ Shaoting Tang,${}^{1}$ Zhiming Zheng,${}^{1}$ Feng Fu,${}^{2,4\ast}$\\
\\
\normalsize{${}^{1}$LMIB, NLSDE, BDBC, PCL and School of Mathematical Sciences,}\\
                     \normalsize{Beihang University, Beijing 100191, China}\\
\normalsize{${}^{2}$Department of Mathematics, Dartmouth College, Hanover, NH 03755, USA}\\
\normalsize{${}^{3}$Department of Sociology, Dartmouth College, Hanover, NH 03755, USA}\\
\normalsize{${}^{4}$Department of Biomedical Data Science, Dartmouth College, Lebanon, NH 03756, USA}\\\normalsize{$^\dagger$Corresponding author. Email: antonio.d.sirianni@dartmouth.edu.}\\
\normalsize{$^\ast$Corresponding author. Email: Feng.Fu@dartmouth.edu.}
}

\date{}

\begin{document} 


\baselineskip24pt

\begin{spacing}{1.0}

\twocolumn[
  \begin{@twocolumnfalse}
  \maketitle 
\begin{sciabstract}

In recent years, social media has increasingly become an important platform for political campaigns, especially elections. It remains elusive how exactly public discourse is driven by the intricate interplay between individual socio-cognitive biases, dueling campaign efforts, and social media platforms. We examine this complex socio-political process by integrating observed retweet networks from the 2016 political networks with an agent-based model of political opinion formation and network structure. Here we show that the range of political viewpoints individuals are willing to consider is a key determinant in the formation of polarized networks and the emergence of echo chambers. We also find that winning majority support in public discourse is determined by both the effort exerted by campaigns and the relative ideological positioning of opposing campaigns. Our results demonstrate how public discourse and political polarization can be modeled as an interactive process of shifting individual opinions, evolving social networks, and political campaigns.

\end{sciabstract}
\end{@twocolumnfalse}
 ]

\section*{Introduction}

People are more likely to accept claims that are coherent with their pre-existing beliefs \cite{plous1993psychology}. Furthermore, people are more likely to seek out individuals with similar beliefs. These two mechanisms, confirmation bias and selective exposure \cite{nickerson1998confirmation,jonas2001confirmation}, can generate polarized and homogeneous clusters of opinion, commonly referred to as “echo chambers” \cite{garrett2009echo}. Echo chamber effects may stop individual from being exposed to information or opinions contrary to their extant beliefs, and further radicalize individuals with extreme beliefs \cite{bessi2015science}. Political polarization in the United States has increased markedly in recent decades \cite{grinberg2019fake}. While confirmation bias and selective exposure are not necessarily new phenomena, new digital environments have arisen that can accelerate and entrench the emergence of echo-chambers.  Furthermore, political campaigns aim to strategically capitalize on these mechanisms and environments for their own benefit. Comprehensively understanding the emergence of echo-chambers and large-scale political polarization in the real world requires a model of collective decision making that integrates individual socio-cognitive biases, the digital ecologies and network structures that facilitate communication, and the strategic efforts of political campaigns. 

Information consumption and diffusion has radically changed along with the rapid development of large-scale social networks \cite{lewis2012social,onnela2010spontaneous}. News outlets are integrated with active online social networks, and can now spread information faster, wider and more effectively \cite{allcott2017social,wang2017promoting}. Furthermore, social network users are no longer just passive recipients of information, but can also create and distribute content, selectively disseminate information, and share their own opinions \cite{kleinberg2013analysis}. These changes have had a substantial impact on public discourse and opinion formation: individuals can easily find evidence that supports their existing ideas and follow those people who hold similar viewpoints, amplifying pre-existing socio-cognitive biases \cite{quattrociocchi2011opinions,ugander2012structural}. Recommendation algorithms further amplify these biases by predicting and delivering the content that individuals are most likely to consume \cite{lazer2015rise, bakshy2015exposure}. 

The seemingly ubiquitous formation of echo chambers on social networks has aroused concern in several different fields \cite{castellano2009statistical,bond201261,liu2020homogeneity}. Sociologists and political scientists are concerned that social polarization caused by politically motivated selective exposure threatens democracies \cite{wardle2017information}. Twitter data shows that political user groups are more likely to retweet users belonging to their own group while supporters of one party rarely interacts with rival party supporters \cite{garimella2018political,colleoni2014echo}. The role of opinion-based confirmation bias in the diffusion of rumors and “fake news” has attracted the attention of network and data scientists \cite{friggeri2014rumor,Wellesley,evans2018opinion,fu2008coevolutionary}, and several empirical studies have focused on social network users’ content consumption patterns in the age of misinformation \cite{zollo2015emotional,webster2012dynamics,bessi2015trend,mocanu2015collective,vosoughi2018spread}. Echo chamber effects can reinforce the diffusion of rumors and intensify the segregation of their believers \cite{schmidt2017anatomy}.

\begin{figure*}[ht]
\centering
\includegraphics[width=0.95\linewidth]{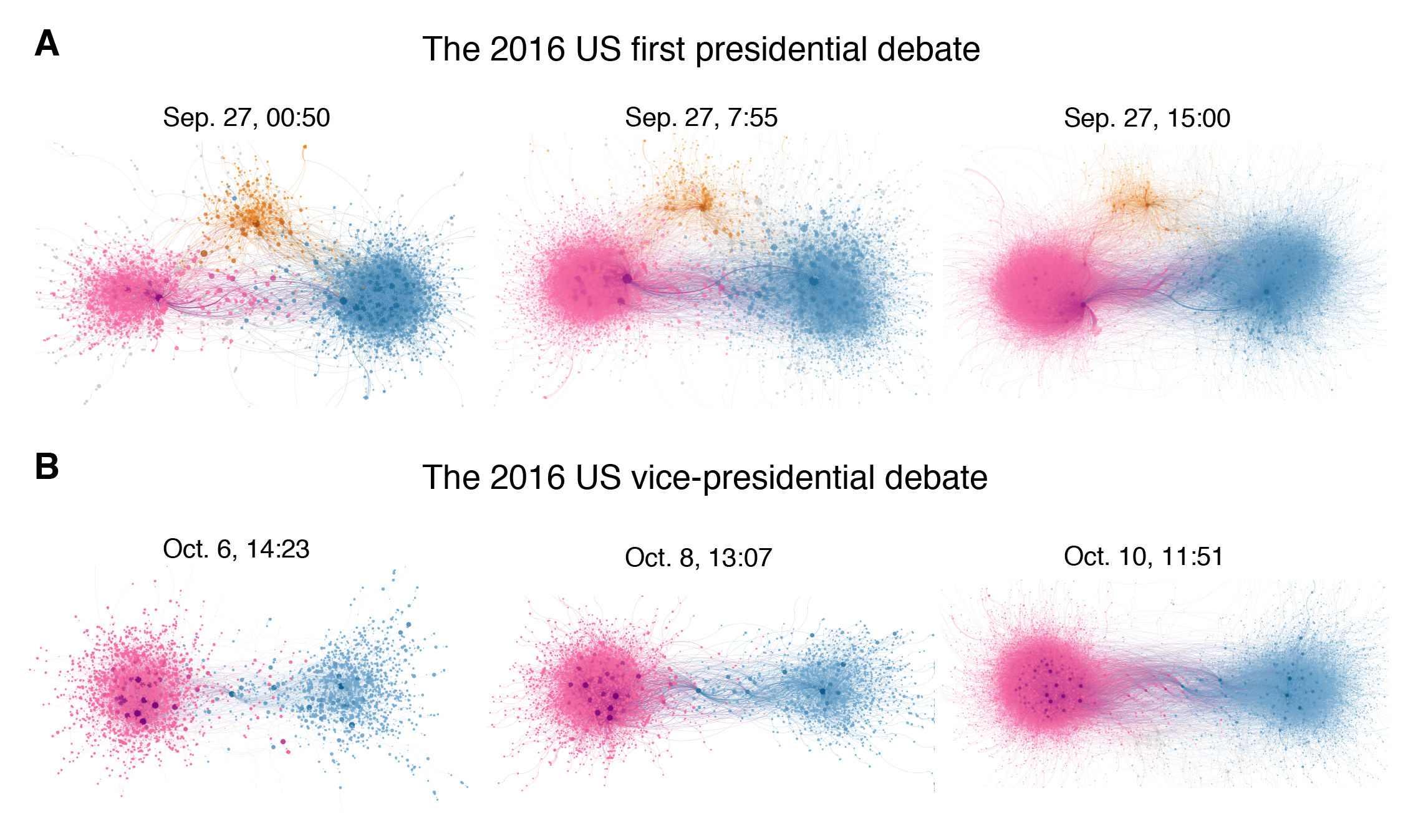}
\caption{ \textbf{Evolving echo-chambers in core retweet networks.} The progression of echo-chamber formation in retweet networks during time frames surrounding (\textbf{A}) the first 2016 U.S. Presidential Debate (\textbf{B})  the 2016 U.S. Vice Presidential Debate. In each subfigure, the nodes represent users, node colors show the node’s cluster, the edges indicate retweets, and node size and color intensity reflect a node’s retweet volume. In (A), the red nodes indicate individuals who tend to support Donald Trump while blue ones Hillary Clinton supporters. The yellow cluster shows users who are mainly retweeting unaligned users or supporters of both sides. The dataset begins at $17:45$ on September $26^{th}$, and the final graph includes $54840$ edges and $25375$ nodes. In (B), the networks are divided into two polarized echo-chambers: the red cluster shows Mike Pence (Trump’s VP) supporters and the blue cluster shows Tim Kaine (Clinton’s VP) supporters. Unlike the first presidential debate, there is not a cluster of undecided supporters, despite both graphs being roughly the same size in terms of nodes and edges. The dataset begins at $15:40$ on October $4^{th}$, $2016$ and the final graph includes $55008$ edges and $24811$ nodes. }
\label{fig1}
\end{figure*}

While there is extensive empirical work documenting instances of echo chamber formation, and simulation-based work that demonstrates the mechanisms sufficient for polarization, few attempts have been made to directly integrate observed instances of polarization with proposed generative mechanisms \cite{deffuant2000mixing,friedkin2016network}. To this end, we develop an agent-based model which explicitly incorporates  socio-cognitive biases and network structure to understand the formation of echo chambers and their influence on the public discourse, and then use this model to estimate the causes of empirically observed instances of polarization. First, we demonstrate the evolution of echo chambers in Twitter retweet networks around the time of a series of major political events (Debates between Candidates in the 2016 U.S. Presidential Elections) \cite{barbera2015tweeting}. We then show that our proposed model of polarization successfully reproduces the observed clustering phenomena. This result suggests that the emergent polarized and segregated network structures are largely a function of the ideological difference between campaigns, and the extent to which individuals are open to considering alternative viewpoints (“open-mindedness”).  We then quantify and explore how confirmation and selection bias known generators of echo chambers \cite{schmidt2017anatomy,del2016spreading}, influence the coevolution of political opinions and network structure. We also find that the success of political campaigns is determined no by relative campaign effort (whose voice is louder) and more subtly by the divergence of ideological positioning between opposing campaigns. Finally, we conduct model-data integration of our proposed model and the observed dynamics of the 2016 US presidential election. Our mechanistic model reproduces the evolution of collective political opinion as measured via Twitter discourse data, providing profound insights into the 2016 campaign. Taken together, our findings demonstrate how shifts in collective opinion can be understood as a function of interacting individual-level cognitive processes, online network structure, and political campaigns. More importantly, these findings suggest potential ways to address political polarization at both the individual and population level, as well as related problems that stem from the social diffusion of information and disinformation.

\section*{Results}


To illustrate the evolution of echo-chambers over time, we display the core Twitter retweet networks during time windows surrounding two highly visible political events in Fig. \ref{fig1}: the first U.S. Presidential Debate in 2016 between Hilary Clinton and Donald Trump, and the lone Vice-Presidential Debate between Tim Kaine and Michael Pence (see Materials and Methods and Fig. S1 for data processing details). During the first US Presidential Debate Twitter users form three distinct clusters: one supporting Clinton (blue), one Trump (red), and a third cluster of seemingly undecided users in the middle (yellow) (Fig. \ref{fig1}A). The middle cluster becomes relatively smaller over time as a higher percentage of users becomes absorbed in the two extremal clusters. The twitter snapshots surrounding the vice-presidential debate 11 days later reflect an even more polarized electorate (Fig. \ref{fig1}B). There are only two main clusters of supporters, one for Kaine (Clinton’s VP candidate) and one for Pence (Trump’s VP candidate), that become more distinct and polarized over time.


\subsection*{Agent-based model of public discourse and echo-chamber formation}

\begin{figure*}[ht]
\centering
\includegraphics[width=0.95\linewidth]{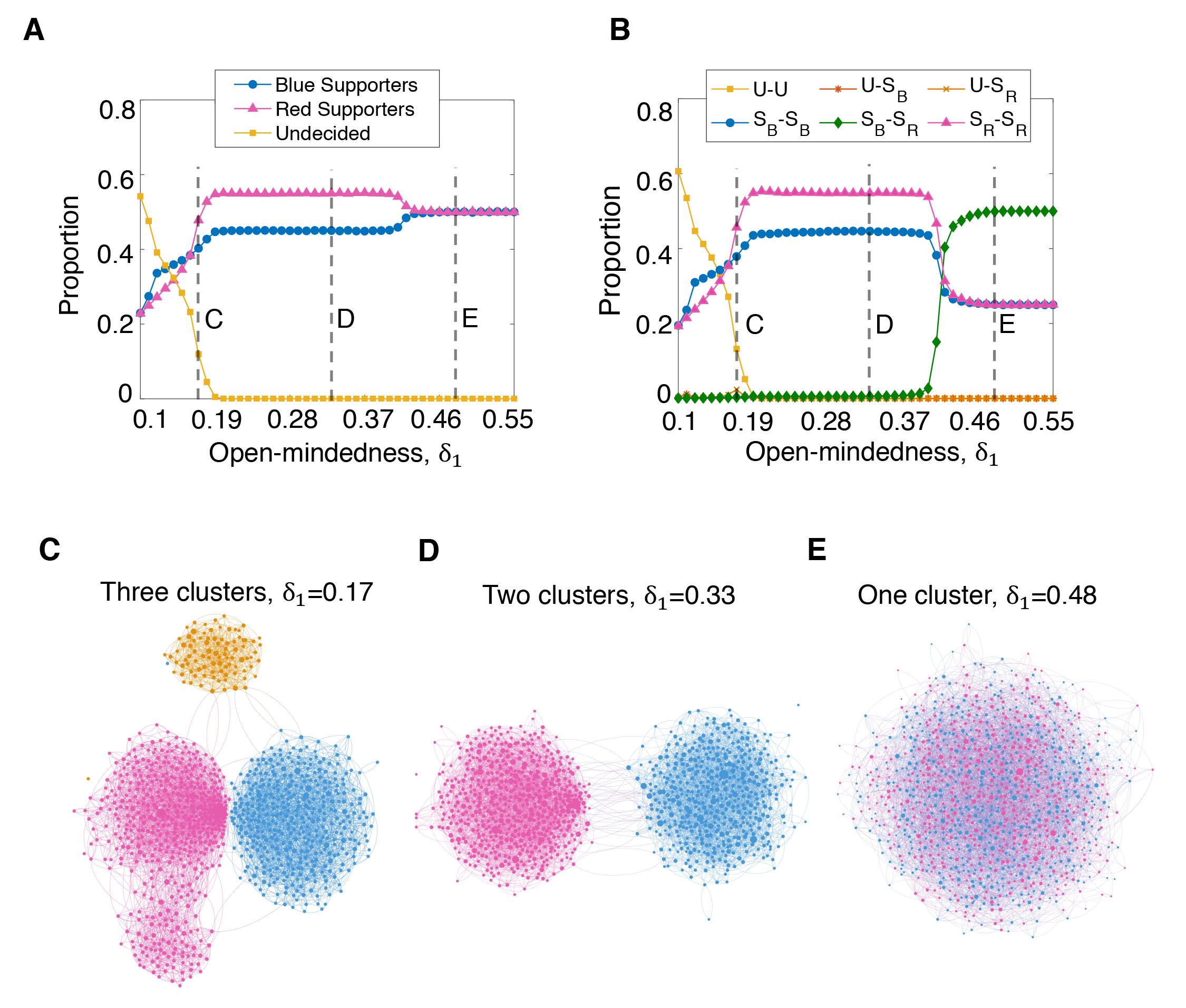}
\caption{ \textbf{Modeling emergence of echo chambers.} The relationship between the degree of ideological segregation and population identity largely influence the nature and outcome of public discourse. Shifts in identity scope are alone sufficient to drastically change the (\textbf{A}) breakdown of candidate support and (\textbf{B}) the distribution of ties between and within supporter groups. The stable network structures of simulations corresponding to three different identity scope values (\textbf{C} to \textbf{E}) show how open-mindedness effects the emergence of echo-chambers. Parameters: campaign effort $\Omega_T=0.5,$ position of viewpoint Blue $\theta_1=0.2,$ position of viewpoint Red $\theta_2=0.7,$ influence of candidate $\mu_1=0.5,$ influence of discussion partner $\mu_2=0.5$. All simulations begin from an ER graph,  the simulations for figures A and B have network size $N=10^4$, and average degree $\langle k \rangle=6$. Simulation results are averaged over $100$ independent runs. In the simulations that generate figures C, D, and E, $N=10^3$ and $\langle k \rangle=6$.}
\label{fig2}
\end{figure*}

We now introduce an agent-based model that can reproduce the patterns of dynamic polarization observed in the core-retweet networks (see Fig. S2 for model parameters), while accounting for the relative effort and position of competing political campaigns. Consider a network $G$ with $N$ users. While several different opinions may be expressed by political and media elites, we assume without loss of generality that there are two: viewpoint Blue ($\theta_1$) and viewpoint Red ($\theta_2$), that each reflect the viewpoint of two opposing political campaigns. Each point exists on ideological spectrum that stretches from $0$ to $1$ ($\theta_i \in [0,1]$), and $\delta_0 = \left|\theta_1-\theta_2\right|$ indicates the degree of ideological separation between the two viewpoints. Each agent, $i$, has an initial belief $w_i \in [0,1]$, that is distributed with uniform probability, and each agent begins as undecided in the race between candidate blue and candidate red. At each time step, an individual will consider updating their support to one of the two candidates, with a probability that reflects the relative effort exerted by each campaign ($\Omega_T$). Their likelihood of accepting updates to their beliefs will depend on their open-mindedness ($\delta_1$), a constant for all individuals in the model, and the proximity of the candidate’s ideology to the individual supporter’s current ideology. Individuals are also connected to others in the population. If a selected agent does not accept a candidate, they will have a conversation with one of their connections and update their beliefs if their connection is ideologically close, and update their connection if the connection is ideologically distant. 

Specifically, the model goes through the following steps at each time step, $T$:
\\[8pt]
$1$. An agent, $i$, is randomly selected.\\
$2$. Agent $i$ considers updating to either $\theta_1$ or $\theta_2$ with probability $\Omega_T$ and $1-\Omega_T$ respectively.\\
$3$. Assume $\theta_1$ is selected. One of three possible interactions occur:

    a. (Support candidate) If $\left|w_i-\theta_1\right|<\delta_1$, then $i$ becomes a Blue supporter. $i$ also becomes more ideologically similar to their new candidate, and $w_i$ is updated to $\bar{w_i}$ beliefs in accordance with a feedback parameter (candidate influence), $\mu_1$: $\bar{w_i}= (1-\mu_1) w_i+ \mu_1 \theta_1$, $\mu_1 \in [0,1]$ \cite{deffuant2000mixing}.
    
    b. (Discuss with friend) If $\left|w_i-\theta_1\right|\ge\delta_1$, then agent $i$ turns has an ideological conversation with $j$. If $\left|w_i-w_j\right|<\delta_1$, $i$ is persuaded by $j$ and updates their beliefs in accordance with a homogeneity parameter (peer influence), $\mu_2$: $\bar{w_i}= (1-\mu_2) w_i+ \mu_2 w_j$, $\mu_2 \in [0,1]$ \cite{lorenz2007continuous}.
    
    c. (Find new friend) If $\left|w_i-\theta_1\right|\ge\delta_1$ and $\left|w_i-w_j\right|\ge\delta_1$ for the chosen candidate and social connection, then $i$ rewires their tie with connection $j$ to another tie, $k$, where $\left|w_i-w_k\right|<\delta_1$.
\\
$4$. Repeat step $1$ to $3$ until all agents update once.
\\

The model then updates to timestep $T+1$, which continues when the beliefs of individual agents have all stabilized, or when the system as a whole reaches a dynamic equilibrium, which occurs when individuals collectively oscillate between supporting different candidates at a constant rate (e.g., Fig. \ref{fig2}E).

This model synthesizes the socio-cognitive biases of the individuals, the evolving structure of the social network, and the relative effort and ideological positions of campaigns. As we shift these parameters from simulation to simulation, we can look at the effect that these interacting mechanisms have on the emergence of distinct ideological clusters (echo-chambers). Of particular interest is how “open-mindedness” can influence this formation of echo-chambers. On one hand, a population of agreeable people may be particularly prone to the influence of candidates, but a population of closed-minded people may re-wire and only listen to those around them.

Fig. \ref{fig2} illustrates how the emergence of echo-chambers is fundamentally related to the open-mindedness of individual voters.  In this series of agent-based models, we hold the two viewpoints of candidates constant ($\theta_1=0.2$, $\theta_2=0.7$), and that campaign efforts exercise equal effort ($\Omega_T = 0.5$). Candidates and connections both have intermediate levels of influence on individual agents ($\mu_1=0.5,$ $\mu_2=0.5$). As the open-mindedness (or identity scope) of individual agents in the population becomes larger, the tendency for clusters and for candidates to support changes. There are two phase shifts that occur at $0.19$, and $0.43$ in this series of models. At low values ($\delta_1<0.19$), individuals tend to form smaller distinct ideological clusters, but these clusters may not all align around a candidate. In the case presented there is only one such cluster, but we can imagine more clusters emerging if more ideological space existed between the two candidates. The stubbornness of individuals protects them from the influence of centralized sources of influence (campaigns), allowing the formation of independent echo chambers.  At medium values ($0.19 < \delta_1 <0.43$) of open-mindedness, all individuals congeal around one of the two candidates, and seek out connections with like-minded people. Each echo-chamber corresponds to a centralized campaign message that individuals will go to correctly, or eventually be dragged into by their increasingly homophilous social networks. Individuals are polarized because they are just open enough to social influence for polarization to occur.  As open-mindedness becomes high $\delta_1>0.43$), however, the population forms one giant cluster. Individuals are willing to consider a wide range of political messages, and initial ideology has a much weaker correlation with the ideology of the candidate that each individual chooses. In these environments, it would be expected that the relative efforts of the campaigns would be very important, as most individuals are open to most options. Individuals in this situation will not rewire their ties, and most political influence will happen between individuals and candidates.  (The complete evolutions of these three network topologies can be found in the Movies S1- S3.)

\subsection*{Peer and candidate influence and ideological alignment}

\begin{figure}[h!]
\centering
\includegraphics[width=\linewidth]{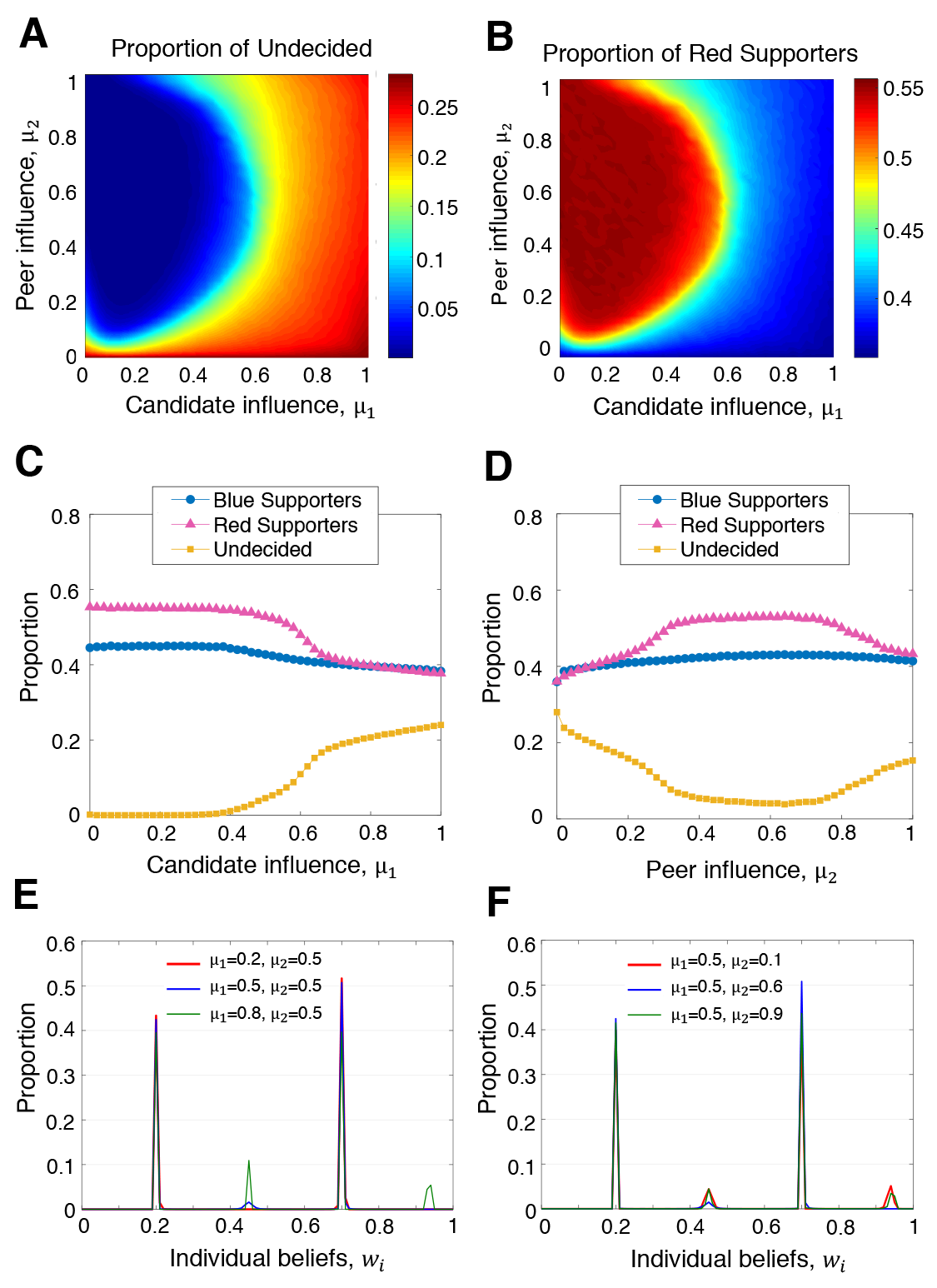}
\caption{ \textbf{Confirmation bias and attitude evolution.} (\textbf{A} to \textbf{B}) Generally, moderate discussion partner influence and weak candidate influence lead to larger echo-chambers (here a significantly higher proportion of Red Supporters) and a lower proportion of undecided individuals. (\textbf{C}) Strong levels of candidate influence may increase the belief distance between supporters and undecided people and accelerate the opinion polarizing process, which hinders sufficient social discourse and results in smaller echo chambers around candidates. (\textbf{D}) When candidate influence is fixed, moderate levels of discussion partner influence promote more social discourse, which reduces the size of the undecided population and effectively prevents the formation of small clusters that hold extreme opinions (also see F). (\textbf{E} to \textbf{F}) Distribution of individual beliefs in steady states, corresponding to the various situations in (C) and (D). Parameters: all simulations begin from an ER graph with $N=10^4$, $\langle k \rangle=6$ and $\Omega_T=0.5,$ $\theta_1=0.2,$ $\theta_2=0.7,$ population open-mindedness $\delta_1=0.18$. In addition, (C) $\mu_2=0.5$, (D) $\mu_1=0.5$. Simulation results are averaged over $100$ independent runs.}
\label{fig3}
\end{figure}

We now turn out attention to how different sources of social influence interact to create echo-chambers or preserve ideological diversity (Fig. \ref{fig3}). In our model there are two parameters that govern social influence processes and the belief shifts of individuals: $\mu_1$ is the extent to which individual beliefs align with those of their selected candidate, $\mu_2$ is the extent to which individual beliefs align with those of their trusted discussion partners.  We would expect high levels of $\mu_1$ to correspond with convergence around candidate viewpoints, and high levels of $\mu_2$ to lead to decentralized and emergent convergence around more arbitrary viewpoints. Intuitively, we would expect that high levels of either or both would lead to echo-chambers and ideological homogeneity.

The results of the model are less intuitive. The emergence of echo-chambers and alignment with political candidates tends to be highest for moderate levels of peer influence, and low levels of candidate influence (Fig. \ref{fig3}A and B). Very high levels of candidate influence will turn all supporters into extremists, and leave candidate supporters beyond the ideological scope of their undecided peers (Fig. \ref{fig3}C, E and Fig. S4 A-C). Low levels of both forms of influence will similarly leave individuals outside the scope of candidates stuck in an undecided part of the ideological spectrum. Extremely high levels of peer influence may similarly leave certain people behind if their peers congeal around viewpoints too quickly. What seems to be most conducive to echo-chamber formation is a moderate process of peer-influence than pulls people into camps slowly without leaving behind undecided stragglers, perhaps guided by very slight top-down influence from candidates (see the phase region where $\mu_1 \in [0, 0.3], \mu_2 \in [0, 0.2]$ in Fig. \ref{fig3}A). High levels of social influence in a population where people are reluctant to hearing different beliefs can create many different ideological islands that people will be “stuck” on, but may prevent the emergence of two dominant and opposing opinions (Fig. \ref{fig3}D, F and Fig. S4 D-E).

\subsection*{The ideological positions and efforts of political campaigns}

\begin{figure*}[ht]
\centering
\includegraphics[width=0.95\linewidth]{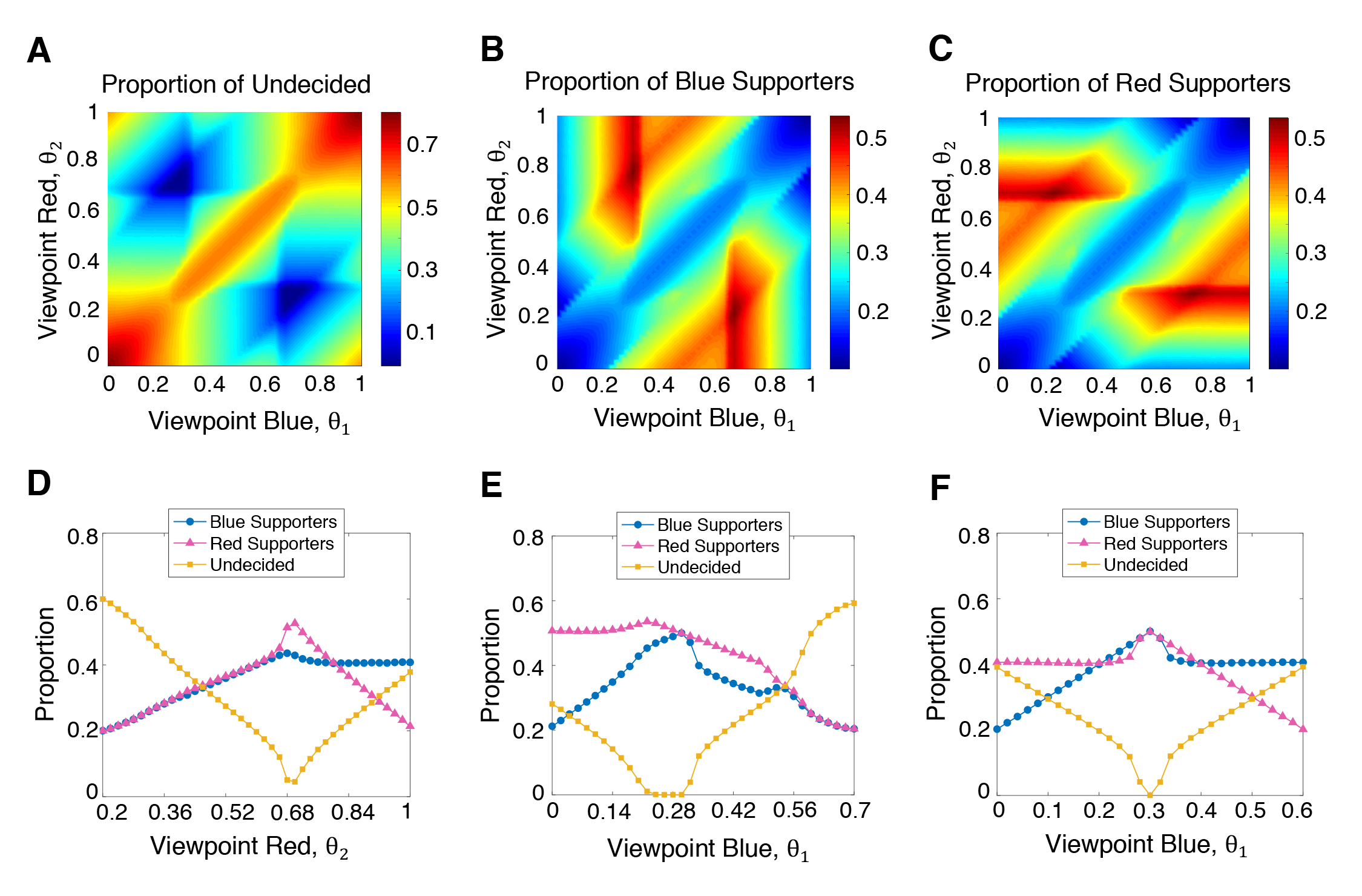}
\caption{ \textbf {Competing campaign ideologies within social networks.} We can use our model to see how campaigns should positions themselves ideologically in order to maximize voter appeal. We assume that in the short run, the open-mindedness (identity scope) of voters and levels of discussion partner and candidate influence on voters will remain constant, so we fix $\delta_1=0.18$, $\mu_1=\mu_2=0.5$. (\textbf{A} to \textbf{C}) show how vote shares for blue, red, and undecided varies as a function of the candidate’s ideological positions. Both candidates maximize vote share by picking opposing position in the ranges of $[0.2,0.3]$ and $[0.7,0.8]$. In (\textbf{D} to \textbf{E}), the ideological campaign of one campaign is fixed ((D) $\theta_1=0.2$ and (E) $\theta_2=0.7$) and the other is varied. In (\textbf{F}), the difference between the two campaigns is settled ($\delta_0=0.4$), and maximum vote share for both candidates occurs when the campaigns are equidistant from $0.5$ ($\theta_1=0.3, \theta_2=0.7$). All simulations begin with an ER graph with $N=10^4$, $\langle k \rangle =6$ and $\Omega_T=0.5$.  Simulation results are averaged over $100$ independent runs.}
\label{fig4}
\end{figure*}

In the presence of social dynamics that lead to the formation of echo-chambers, it may also be logical for political candidates may take on positions that are more extreme than is predicted by standard “median-voter” models of political competition \cite{hotelling1990stability, downs1957economic}. Given a fixed level of the identity scope of individual voters, peer influence, and candidate influence, candidates may be incentivized to take on viewpoints that are more opposed to one another, as opposed to converging around the median. In Fig. \ref{fig4}A-C, we can see that the best response of a candidate is not to simply position themselves immediately to the side of their opponent that is closer to the median, but rather to stake out a position that is positioned in a large gap between their opponent and an end of the political spectrum. 
Vote share by viewpoint for different candidates in different scenarios are displayed in Figures 4D-4F, we see that Red’s best response to a Blue viewpoint of $0.2$ is around $0.68$ (Fig. \ref{fig4}A), Blue’s best response to a Red viewpoint of $0.7$ is around $0.3$ (Fig. \ref{fig4}E), and that given a fixed viewpoint distance between Red and Blue of $0.4$, both candidates maximize proportion of their support by aligning symmetrically around the median ($0.5$) and selecting viewpoints of $0.3$ and $0.7$.

\begin{figure}[h!]
\centering
\includegraphics[width=\linewidth]{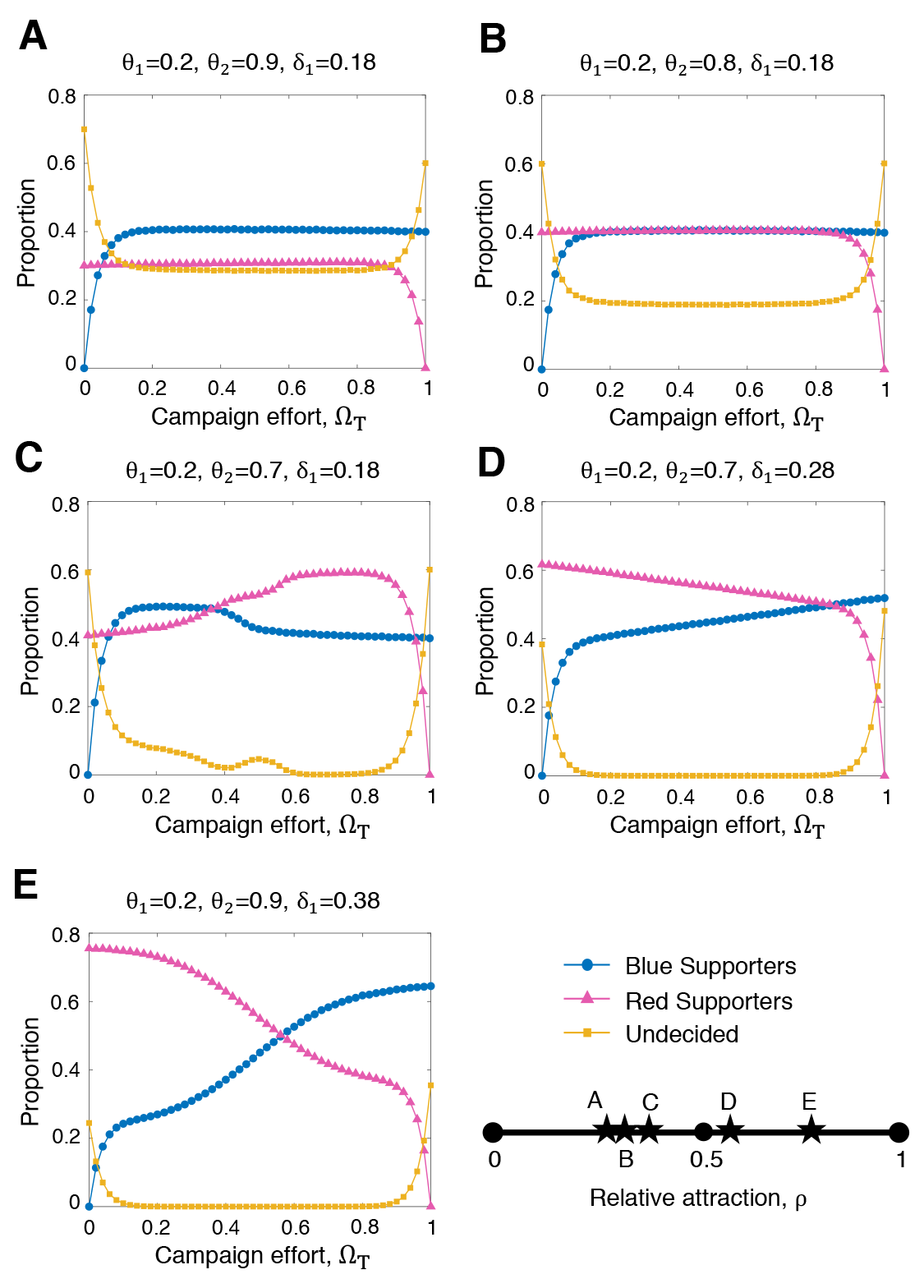}
\caption{ \textbf{Impact of campaign effort on winning majority support.} (\textbf{A} to \textbf{E}) As the relative attraction of the viewpoints (defined as identity scope divided by candidate spread) in campaign competition increases ($\rho=0.257, 0.3, 0.36, 0.56, 0.76$, respectively), the relative efforts of campaigns become more important. In some cases, the effects of campaign effort become counterproductive (see (C)). Parameters: all simulations begin form an ER graph with $N=10^4$, $\langle k \rangle =6$ and $\mu_1=\mu_2=0.5$. Simulation results are averaged over $100$ independent runs.}
\label{fig5}
\end{figure}

We also study the influence of campaign effort on winning majority support (Fig. \ref{fig5}). The importance of campaign effort in our model is largely dependent on the ratio of the identity scope ($\delta_1$) of individuals to the spread of the candidate positions ($\delta_0$). We denote this value as $\rho={\delta_1}/{\delta_0}$. When $\rho=1/2$, by definition there are no voters between the two candidates who could possibly avoid the influence of either candidate, but in the absence of peer influence each will fall in line with the candidate who is closest to them. (Fig. \ref{fig5}) shows how campaign effort varies in effectiveness as a function of $\rho$.  As $\rho$ goes from $0$ to $1/2$, the effects of relative campaign efforts may be trivial in the absence of large differences (Fig. \ref{fig5}A, B). For levels of $\rho$ slightly lower than $1/2$, excessive campaigning may even be counter-productive if campaigns accidentally “pull-in” potential supporters too quickly, leaving undecideds more vulnerable to their opponent (Fig. \ref{fig5}C). As $\rho$ goes from $1/2$ to $1$, however, the number of voters who could support either candidate increases, increasing the importance of campaign efforts (Fig. \ref{fig5}D, E). We also see across all models that there are much higher numbers of “undecideds” if campaign efforts are lopsided, as the message of one candidate is effectively drowned out by their opponent.

Taken together, our results show that the social conditions that lead to echo-chambers can create political incentives that oppose our common-sense notions of how to win an election. Campaigns no longer benefit from converging to the center, as the median voter theorem predicts. Furthermore, in certain cases where candidates are not too far together or too far apart ideologically, campaign efforts can be counter-productive, as they may polarize individuals that would otherwise be useful in winning over moderates and undecideds. This is an important insight, especially given the relative spending and outcome of the 2016 U.S. Presidential Election.

\subsection*{Model-data integration of opinion evolution in the 2016 US presidential election}

\begin{figure*}[htbp]
\centering
\includegraphics[width=0.95\linewidth]{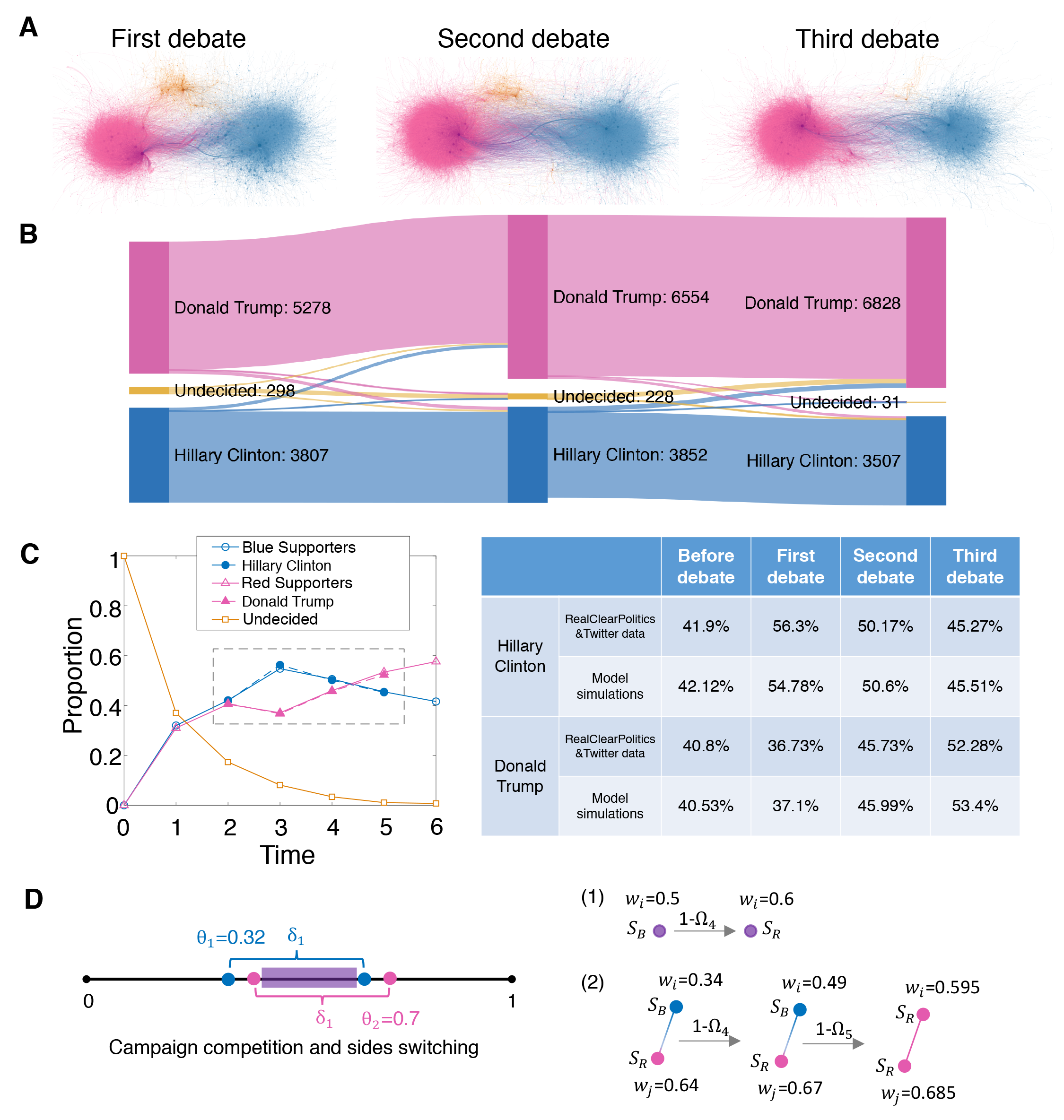}
\caption{ \textbf{Model-data integration provides insights into the 2016 US presidential election.} (\textbf{A}) Retweet networks of the first, second, and third presidential debate in 2016 US presidential election show the rise of echo chambers that gradually absorb undecided voters. (\textbf{B}) Sankey diagram of individual-level opinion evolutions for Twitter users between the first and second debates, and the second and third debates. Most Donald Trump or Hillary Clinton supporters are ``loyal followers” who would not change their political tendency from beginning to end. (\textbf{C}) Reproduce opinion evolution processes in the 2016 US presidential election using an agent-based model. We take the Real Clear Politics polling average on September 15, 2016 as an initial estimate of pre-debate support, while from the first to third debates, we use the same Twitter data as in (A). Fixing a middle level of confirmation bias ($\mu_1= \mu_2=0.5$), we find a group of best fitting parameters for the ideological positions of campaigns, the identity scope (open-mindedness) of voters, and relative levels of campaign effort: $\theta_1=0.32,$ $ \theta_2=0.7,$ $\delta_1=0.32,$ $\Omega_1= \Omega_2=0.5,$ $\Omega_3=0.8,$ $ \Omega_4=0.4,$ $\Omega_5=0.1$ (We let $\Omega_6=\Omega_5=0.1$ to model one extra time-step). Each result in the figure is the average of $100$ simulations. (\textbf{D}) Illustration of the potential mechanisms that lead to side switching within this model (in this case from Blue supporter to Red supporter). Two possible ways are shown: (1) a blue supporter within ideological “range” of both candidates switches via one single update as a result of campaigning; (2) switch via multiple updates, a voter is not persuaded by the candidate initially, but instead talks to a discussion partner within their range who is a red-supporter, and then updates their beliefs. In the next time step the voter is reached by the red campaign again, and is close enough to the red candidate to be persuaded.} 
\label{fig6}
\end{figure*}

Finally, we integrate our model with data from the 2016 U.S. Presidential Campaign (Fig. \ref{fig6}). As the campaign unfolds, data from Twitter shows that network polarization increases as echo-chambers absorb undecided voters over time (Fig. \ref{fig6}A). Looking at the trajectories of individual Twitter users retweet patterns (Fig. \ref{fig6}B), we see that most individuals who initially supported Clinton or Trump remained fairly constant over time, with Donald Trump winning over many undecideds between the second and third debates.. 

Using the proposed model in tandem with observed data, we can estimate the ideological position of both candidates, the population identity scope, as well as how the relative efforts and efficacy of each campaign shifted in the last few months of the U.S. Presidential campaign, while holding levels of confirmation bias within voters constant. Fig. \ref{fig6}C shows how our agent-based model of opinion evolution aligned with shifting levels of support for Clinton and Trump during the 2016 campaign. We assume that the ideological viewpoints of both campaigns ($\theta_1$, $\theta_2$) as well as the population identity scope ($\delta_1$) are unchanged throughout the campaign. Simulations start from an Erdos-Renyi graph with $N=10^4$ and $\langle k \rangle=4.5$, which replicates the size and degree distribution of the observed retweet data. We set a moderate level of population confirmation bias ($\mu_1=\mu_2=0.5$). We then find a group of best fitting parameters for the ideological polarization of the candidates and the level of campaign effort at each of the time steps ($\theta_1=0.32,$ $ \theta_2=0.7,$ $\delta_1=0.32,$ $\Omega_1= \Omega_2=0.5,$ $\Omega_3=0.8,$ $ \Omega_4=0.4,$ $\Omega_5=0.1$) using least squares minimization with a simulated annealing algorithm (see Materials and Methods). 

These results reveal just how dynamic and unpredictable the final phases of the 2016 U.S. Presidential Campaign were, with Clinton and Trump rapidly trading advantages in relative campaigns. Both candidates had ideological positions that were estimated to be a moderate distance from the median of $0.5$, with the Clinton campaign being slightly closer ($\theta_1=0.32,$ $ \theta_2=0.7$). Furthermore, the large estimated identity scope of $0.32$ places a large proportion of voters within ideological reach of both campaigns. The model also shows wide swings in the relative effort of both campaign, reflecting a campaign where high levels of spending and social media outreach, news reports, and debate performances could rapidly change each candidates’ appeal to undecided voters.

The observed data also shows that a small fraction of individuals may have switched camps during the election (Fig. \ref{fig6}B). Fig. \ref{fig6}D illustrates how this process could occur within our proposed model – a ``blue” supporter with neutral beliefs that are within reach of both camps (left side of Fig. \ref{fig6}D), may be directly persuaded by the ``red” campaign (Fig. \ref{fig6}D(1)). Or, a supporter with ``blue” beliefs may interact with a ``red” neighbor and be brought to a spot within reach of the “red” campaign (Fig. \ref{fig6}D(2)). Overall, this is an example of how open-mindedness can increase the importance of campaign efforts. While polarization and voter stubbornness in the face of new information is detested in modern politics, in these examples it makes individuals immune from falling under the influence of high-spending campaigns.

\section*{Discussion}

Several empirical studies from large-scale social networks have provided valuable insights into the emergence of echo chambers. However, comprehensive analysis of collective opinion dynamics requires simultaneous consideration of top-down campaign level decisions, the network structure of political communication, interpersonal influence dynamics, and individual cognitive biases. Our analysis offers a theoretically driven agent-based model that can estimate and account for the effect of these mechanisms using real-world data. 

Interactions between human behavior and social systems are often non-linear and complex, and the aggregate consequences of small changes in individual behavior can be counterintuitive. The model we present here is no exception. A surprising insight is that bipartisan echo-chambers may result from moderate levels of open-mindedness, where people are susceptible to slightly different beliefs but not overwhelmingly different beliefs, as opposed to extreme closed-mindedness (which preserves ideological diversity) or extreme open-mindedness (which brings everyone together somewhere in the middle). This is an inherently uncomfortable finding. The idea of voters being ideologically flexible in the face of new facts and opinions and arguments, while still maintaining a set of core principles, sounds virtuous and essential to a functioning democracy. Yet this `bend but don’t break’ style of opinion updating may be the very thing that leads to polarization. A purely stubborn or conforming population may both be less susceptible to the emergence of two dominant and opposing echo-chambers.

While the formation of echo-chambers and the issue of political polarization is the primary focus of our model, our results also have practical implications for political campaign strategy. Our findings suggest a potential downside to aggressive campaigning from persuasive but extreme candidates: supporters may become too radical for their relatively moderate friends. In some cases, campaigns and candidates may be better off by being less persuasive, as this gives their supporters ideological room to make their candidate’s case in everyday discourse. Candidates with relatively extremal positions may do well by instead adopting a ``big tent” strategy where they win supporters without persuading them to change their ideological stance. In the age of social media, political strategists may need to think more about the order in which they’re targeting people, and their commitment to ideological positions. While ``firing up the base” may be important early on for gathering donations, it may also create a rift between supporters and potential converts. Preaching to the choir may make it difficult to fill up the pews.

Social division along political lines has aroused great concern since the 2016 US presidential election \cite{allcott2017social,kucharski2016post,bovet2019influence}. But the emergence of echo-chambers is not confined to elections. In the age of ``fake news”, determining the truth in everyday affairs has also become a collective decision-making process prone to interacting cognitive biases and social influence dynamics. The spread of misinformation on social networks threatens the foundations of democracy and often disrupts everyday life \cite{pennycook2019fighting,williamson2016take,lazer2018science}. The insights generated from our model suggest not only several techniques for inhibiting the emergence of echo-chambers, but also for how to develop more effective messaging strategies for import public health issues, such as the importance of vaccinations or the potential dangers of vaping. In these cases, the public has a strong interest in generating a single echo-chamber inhabited by the right idea. Our models suggest that this would be best accomplished through moderate messaging that gradually pulls individuals with extreme anti-vaccine or pro-vaping beliefs closer to the mainstream, as opposed to stronger messaging that may convert moderate skeptics but leave extremists isolated in a smaller echo-chamber. Ultimately, what turns a set of like-minded individuals into an ideological bubble is the disappearance of other people who can come in and change their mind.


\section*{Materials and Methods}

\subsection*{Twitter data processing}

All original Twitter datasets are publicly available at \emph{https://www.docnow.io/catalog}. The raw twitter data are not allowed to be shared online according to Twitter's terms of service, therefore all the datasets are Twitter IDs. We first download the tweets using Twitter's API and obtain user ID (nodes), retweeting relationship (edges) and the timestamp from each tweet. Computational constraints prevents us from plotting the whole network, so we use a random sample to obtain $10$-$20\%$ data of the original datasets for the first, second and third debate, and then focus on the largest connected component (which contains over $90\%$ of nodes). Layouts are constructed using the \emph{Force Atlas 2} algorithm in \emph{Gephi}. Nodes are weighted by the number of times they are retweeted. To capture the core structural characteristics of opinion evolution without losing any information, we record all nodes’ weights and get rid of “pure followers” who are not being retweeted (in-degree equals to $0$) and only retweet others once (out-degree equals to $1$). This leads to a more simplified retweet network with weighted nodes (Fig. S1). This simplification takes out about half of all edges and improves the precision of the community analysis. The retweet networks in Fig. 6A include $54840$ edges and $25375$ nodes for the first debate, $81055$ edges and $33995$ nodes for the second debate, and $58354$ edges and $26219$ nodes for the third debate. Finally, the weighted supporter estimates for each camp (or the supporter strength observed on social networks) can be approximately calculated by simply summing the weight of all nodes within each echo chamber. The sample datasets we use in this paper and the data processing codes (written in Python 3) can be obtained at  \emph{https://github.com/fufeng/Public-discourse-and-social-network-echo-chambers}.

All the network figures in this paper use the \emph{Force Atlas 2} layout algorithm from \emph{Gephi} \cite{jacomy2014forceatlas2}. The color for each node is calculated using the \emph{Louvain Algorithm} for community detection \cite{blondel2008fast}.

We stress that our main focus is to understand how public opinion evolves through individual decision-making processes and socio-cognitive biases using agent-based models. Small inaccuracies in the true proportions of candidate supporters that are caused by random sampling and the community detection algorithm are unavoidable but tolerable for our purposes.

\subsection*{Data fitting method}

Within the time-scope of a single event, such as the 2016 US presidential election, we assume that the inherent psychological and cognitive properties (i.e., the identity scope and the effect of discussion partner and candidate influence) of Twitter users would not change by a discernable amount. Therefore, we aimed to find a static set of values for $\theta_1$, $\theta_2$, $\delta_1$, $\mu_1$, $\mu_2$ with a changing group of $\Omega_T$ that can best mimic the evolution of the campaign. The simulation results of four successive time steps (from $T=2$ to $T=5$) are expected to reproduce the real data. To reduce the computational complexity of the fitting procedure we fix a moderate degree of candidate influence and discussion partner influence and set $\mu_1=\mu_2=0.5$. Other parameters are constrained as follows: $\theta_1\in [0,0.5]$, $\theta_2\in [0.5,1]$, $\delta_1 \in [0,\theta_2-\theta_1]$, and $\Omega_T\in [0,1]$, $T\in\{1,2,3,4,5\}$. The step length for $\theta_1$, $\theta_2$ and $\delta_1$ is $0.02$ while the step length for $\Omega_T$ is $0.1$. We take the average value of $100$ times simulations as the final results for each group of parameters. We develop a simple simulated annealing (SA) algorithm with least squares minimization to find the best fit of real data, which is as follows:

1. Initialization and definitions. Define $S=\{\theta_1, \theta_2, \mu_1, \Omega_T \}$ as the set of all possible combinations of parameters. According the descriptions of parameter range and step length above, $S$ is a finite set. Denote $R$ as the eight-dimensional vector of real data (containing the proportions of supporters for two candidates before debate and during the first, second and third debates). Initially, the iteration counter $i=0$ and an initial state is set $S_0 \in S$. In addition, we select a starting temperature $T_0$ and a freezing threshold $T_{min}$ with a cooling rate $r$. At time step $i$, $S_i$ is the current state and $T_i$ is the cooling schedule. Based on the least squares minimization, we define the cost function $f(S_i)=\frac{1}{N} \cdot \frac{\lVert Aver-R \rVert ^2}{\sigma^2}$, where $N=8$ is the dimension of $R$, $Aver$ is the average result (also a eight-dimensional vector) of $100$ simulations for state $S_i$, $\lVert Aver-R \rVert$ is the Euclidean distance of vector $Aver$ and $R$, and $\sigma=1\%$ is a precision control coefficient for the global fitting.

2. The SA methods. For time step $i+1$, a new state $\bar S$ is randomly generated from $S-\{S_i\}$, and the Markov chain of the state vector $S_i$ is determined by the followings: 

\begin{algorithm}
\caption{Simulated Annealing}
\While {$T_i >T_{min} \wedge f(S_i)\geq1$}{
      \eIf {$f(\bar S) \leq f(S_i)$} {
       $S_{i+1}=\bar S$\;
       }{
       $S_{i+1}=\bar S$ with probability $\exp \left[(f(S_i)-f(\bar S))/T_i \right]$\;
       $S_{i+1}=S_i$ otherwise.
       }
    $T_{i+1}=r*T_i$
}
\end{algorithm}

The algorithm allows prior information and can be recycled by setting $S_0$ as the best group of parameters that we have already found. The C++ code for this simulated annealing algorithm along with our dynamic model is available at \emph{https://github.com/fufeng/Public-discourse-and-social-network-echo-chambers}.

\section*{Supplementary Materials} 
Supplementary Text \\
Figs. S1-S9 \\
Movies S1-S3: time evolutions of the three network topologies shown in Fig. 2C-E.

\noindent \textbf{Acknowledgements:} X.W., AD.S., S.T., Z.Z. \& F.F. gratefully acknowledge David Rand who provides constructive suggestions on examining another dynamical model (as we provided in Supplementary Materials) and adding individual-level observations in the 2016 US presidential election.\\
\noindent \textbf{Funding:} This work is supported by Program of National Natural Science Foundation of China Grant No. 11871004, 11922102. F.F. is supported by a Junior Faculty Fellowship awarded by the Dean of the Faculty at Dartmouth and also by the Bill \& Melinda Gates Foundation (award no. OPP1217336), the NIH COBRE Program (grant no.1P20GM130454), the Neukom CompX Faculty Grant, the Dartmouth Faculty Startup Fund, and the Walter \& Constance Burke Research Initiation Award.\\
\noindent \textbf{Author Contributions:} X.W. AD.S. and F.F. conceived the project, X.W., F.F. carried out the models, X.W, S.T. and Z.Z. analyze the data and conduct the simulations, X.W., AD.S. and F.F. wrote the manuscript.\\
\noindent \textbf{Competing Interests:} The authors declare that they have no competing financial interests.\\
\noindent \textbf{Data and materials availability:} All original Twitter datasets are open-sourced and are available on the website \url{https://www.docnow.io/catalog/}. The randomly sampled data we use in this paper, the data processing code using Python 3 environment, and the C++ code for simulated annealing algorithm along with the agent-based model in the main text are all available at \url{https://github.com/fufeng/Public-discourse-and-social-network-echo-chambers}. The poll data before debate can be found at \url{https://www.realclearpolitics.com/epolls/2016/president/us/general_election_trump_vs_clinton_vs_johnson_vs_stein-5952.html}. 
\end{spacing}

\clearpage


\begin{thebibliography}{10}

\bibitem{plous1993psychology}
S.~Plous, {\it The psychology of judgment and decision making.\/} (Mcgraw-Hill
  Book Company, 1993).

\bibitem{nickerson1998confirmation}
R.~S. Nickerson, Confirmation bias: A ubiquitous phenomenon in many guises.
\newblock {\it Review of general psychology\/} {\bf 2}, 175 (1998).

\bibitem{jonas2001confirmation}
E.~Jonas, S.~Schulz-Hardt, D.~Frey, N.~Thelen, Confirmation bias in sequential
  information search after preliminary decisions: an expansion of dissonance
  theoretical research on selective exposure to information.
\newblock {\it Journal of personality and social psychology\/} {\bf 80}, 557
  (2001).

\bibitem{garrett2009echo}
R.~K. Garrett, Echo chambers online?: Politically motivated selective exposure
  among internet news users.
\newblock {\it Journal of Computer-Mediated Communication\/} {\bf 14}, 265--285
  (2009).

\bibitem{bessi2015science}
A.~Bessi, M.~Coletto, G.~A. Davidescu, A.~Scala, G.~Caldarelli,
  W.~Quattrociocchi, Science vs conspiracy: Collective narratives in the age of
  misinformation.
\newblock {\it PloS one\/} {\bf 10}, e0118093 (2015).

\bibitem{grinberg2019fake}
N.~Grinberg, K.~Joseph, L.~Friedland, B.~Swire-Thompson, D.~Lazer, Fake news on
  twitter during the 2016 us presidential election.
\newblock {\it Science\/} {\bf 363}, 374--378 (2019).

\bibitem{lewis2012social}
K.~Lewis, M.~Gonzalez, J.~Kaufman, Social selection and peer influence in an
  online social network.
\newblock {\it Proceedings of the National Academy of Sciences\/} {\bf 109},
  68--72 (2012).

\bibitem{onnela2010spontaneous}
J.-P. Onnela, F.~Reed-Tsochas, Spontaneous emergence of social influence in
  online systems.
\newblock {\it Proceedings of the National Academy of Sciences\/} {\bf 107},
  18375--18380 (2010).

\bibitem{allcott2017social}
H.~Allcott, M.~Gentzkow, Social media and fake news in the 2016 election.
\newblock {\it Journal of Economic Perspectives\/} {\bf 31}, 211--36 (2017).

\bibitem{wang2017promoting}
X.~Wang, W.~Li, L.~Liu, S.~Pei, S.~Tang, Z.~Zheng, Promoting information
  diffusion through interlayer recovery processes in multiplex networks.
\newblock {\it Physical Review E\/} {\bf 96}, 032304 (2017).

\bibitem{kleinberg2013analysis}
J.~Kleinberg, Analysis of large-scale social and information networks.
\newblock {\it Phil. Trans. R. Soc. A\/} {\bf 371}, 20120378 (2013).

\bibitem{quattrociocchi2011opinions}
W.~Quattrociocchi, R.~Conte, E.~Lodi, Opinions manipulation: Media, power and
  gossip.
\newblock {\it Advances in Complex Systems\/} {\bf 14}, 567--586 (2011).

\bibitem{ugander2012structural}
J.~Ugander, L.~Backstrom, C.~Marlow, J.~Kleinberg, Structural diversity in
  social contagion.
\newblock {\it Proceedings of the National Academy of Sciences\/} p. 201116502
  (2012).

\bibitem{lazer2015rise}
D.~Lazer, The rise of the social algorithm.
\newblock {\it Science\/} {\bf 348}, 1090--1091 (2015).

\bibitem{bakshy2015exposure}
E.~Bakshy, S.~Messing, L.~A. Adamic, Exposure to ideologically diverse news and
  opinion on facebook.
\newblock {\it Science\/} {\bf 348}, 1130--1132 (2015).

\bibitem{castellano2009statistical}
C.~Castellano, S.~Fortunato, V.~Loreto, Statistical physics of social dynamics.
\newblock {\it Reviews of modern physics\/} {\bf 81}, 591 (2009).

\bibitem{bond201261}
R.~M. Bond, C.~J. Fariss, J.~J. Jones, A.~D. Kramer, C.~Marlow, J.~E. Settle,
  J.~H. Fowler, A 61-million-person experiment in social influence and
  political mobilization.
\newblock {\it Nature\/} {\bf 489}, 295 (2012).

\bibitem{liu2020homogeneity}
L.~Liu, X.~Wang, Y.~Zheng, W.~Fang, S.~Tang, Z.~Zheng, Homogeneity trend on
  social networks changes evolutionary advantage in competitive information
  diffusion.
\newblock {\it New Journal of Physics\/} {\bf 22}, 013019 (2020).

\bibitem{wardle2017information}
C.~Wardle, H.~Derakhshan, Information disorder: Toward an interdisciplinary
  framework for research and policymaking.
\newblock {\it Council of Europe report, DGI (2017)\/} {\bf 9} (2017).

\bibitem{garimella2018political}
K.~Garimella, G.~De~Francisci~Morales, A.~Gionis, M.~Mathioudakis, {\it
  Proceedings of the 2018 World Wide Web Conference on World Wide Web\/}
  (International World Wide Web Conferences Steering Committee, 2018), pp.
  913--922.

\bibitem{colleoni2014echo}
E.~Colleoni, A.~Rozza, A.~Arvidsson, Echo chamber or public sphere? predicting
  political orientation and measuring political homophily in twitter using big
  data.
\newblock {\it Journal of Communication\/} {\bf 64}, 317--332 (2014).

\bibitem{friggeri2014rumor}
A.~Friggeri, L.~Adamic, D.~Eckles, J.~Cheng, {\it Eighth International AAAI
  Conference on Weblogs and Social Media\/} (2014).

\bibitem{Wellesley}
P.~Metaxas, Evidence of pizzagate conspiracy theory on twittertrails,
  \url{https://blogs.wellesley.edu/twittertrails/}.

\bibitem{evans2018opinion}
T.~Evans, F.~Fu, Opinion formation on dynamic networks: identifying conditions
  for the emergence of partisan echo chambers.
\newblock {\it Royal Society open science\/} {\bf 5}, 181122 (2018).

\bibitem{fu2008coevolutionary}
F.~Fu, L.~Wang, Coevolutionary dynamics of opinions and networks: From
  diversity to uniformity.
\newblock {\it Physical Review E\/} {\bf 78}, 016104 (2008).

\bibitem{zollo2015emotional}
F.~Zollo, P.~K. Novak, M.~Del~Vicario, A.~Bessi, I.~Mozeti{\v{c}}, A.~Scala,
  G.~Caldarelli, W.~Quattrociocchi, Emotional dynamics in the age of
  misinformation.
\newblock {\it PloS one\/} {\bf 10}, e0138740 (2015).

\bibitem{webster2012dynamics}
J.~G. Webster, T.~B. Ksiazek, The dynamics of audience fragmentation: Public
  attention in an age of digital media.
\newblock {\it Journal of communication\/} {\bf 62}, 39--56 (2012).

\bibitem{bessi2015trend}
A.~Bessi, F.~Zollo, M.~Del~Vicario, A.~Scala, G.~Caldarelli, W.~Quattrociocchi,
  Trend of narratives in the age of misinformation.
\newblock {\it PloS one\/} {\bf 10}, e0134641 (2015).

\bibitem{mocanu2015collective}
D.~Mocanu, L.~Rossi, Q.~Zhang, M.~Karsai, W.~Quattrociocchi, Collective
  attention in the age of (mis) information.
\newblock {\it Computers in Human Behavior\/} {\bf 51}, 1198--1204 (2015).

\bibitem{vosoughi2018spread}
S.~Vosoughi, D.~Roy, S.~Aral, The spread of true and false news online.
\newblock {\it Science\/} {\bf 359}, 1146--1151 (2018).

\bibitem{schmidt2017anatomy}
A.~L. Schmidt, F.~Zollo, M.~Del~Vicario, A.~Bessi, A.~Scala, G.~Caldarelli,
  H.~E. Stanley, W.~Quattrociocchi, Anatomy of news consumption on facebook.
\newblock {\it Proceedings of the National Academy of Sciences\/} {\bf 114},
  3035--3039 (2017).

\bibitem{deffuant2000mixing}
G.~Deffuant, D.~Neau, F.~Amblard, G.~Weisbuch, Mixing beliefs among interacting
  agents.
\newblock {\it Advances in Complex Systems\/} {\bf 3}, 87--98 (2000).

\bibitem{friedkin2016network}
N.~E. Friedkin, A.~V. Proskurnikov, R.~Tempo, S.~E. Parsegov, Network science
  on belief system dynamics under logic constraints.
\newblock {\it Science\/} {\bf 354}, 321--326 (2016).

\bibitem{barbera2015tweeting}
P.~Barber{\'a}, J.~T. Jost, J.~Nagler, J.~A. Tucker, R.~Bonneau, Tweeting from
  left to right: Is online political communication more than an echo chamber?
\newblock {\it Psychological science\/} {\bf 26}, 1531--1542 (2015).

\bibitem{del2016spreading}
M.~Del~Vicario, A.~Bessi, F.~Zollo, F.~Petroni, A.~Scala, G.~Caldarelli, H.~E.
  Stanley, W.~Quattrociocchi, The spreading of misinformation online.
\newblock {\it Proceedings of the National Academy of Sciences\/} {\bf 113},
  554--559 (2016).

\bibitem{lorenz2007continuous}
J.~Lorenz, Continuous opinion dynamics under bounded confidence: A survey.
\newblock {\it International Journal of Modern Physics C\/} {\bf 18},
  1819--1838 (2007).

\bibitem{hotelling1990stability}
H.~Hotelling, {\it The Collected Economics Articles of Harold Hotelling\/}
  (Springer, 1990), pp. 50--63.

\bibitem{downs1957economic}
A.~Downs, {\it et~al.\/}, An economic theory of democracy  (1957).

\bibitem{williams2018digital}
C.~B. Williams, G.~J. Gulati, Digital advertising expenditures in the 2016
  presidential election.
\newblock {\it Social Science Computer Review\/} {\bf 36}, 406--421 (2018).

\bibitem{kucharski2016post}
A.~Kucharski, Post-truth: Study epidemiology of fake news.
\newblock {\it Nature\/} {\bf 540}, 525 (2016).

\bibitem{bovet2019influence}
A.~Bovet, H.~A. Makse, Influence of fake news in twitter during the 2016 us
  presidential election.
\newblock {\it Nature communications\/} {\bf 10}, 7 (2019).

\bibitem{pennycook2019fighting}
G.~Pennycook, D.~G. Rand, Fighting misinformation on social media using
  crowdsourced judgments of news source quality.
\newblock {\it Proceedings of the National Academy of Sciences\/} p. 201806781
  (2019).

\bibitem{williamson2016take}
P.~Williamson, Take the time and effort to correct misinformation.
\newblock {\it Nature News\/} {\bf 540}, 171 (2016).

\bibitem{lazer2018science}
D.~M. Lazer, M.~A. Baum, Y.~Benkler, A.~J. Berinsky, K.~M. Greenhill,
  F.~Menczer, M.~J. Metzger, B.~Nyhan, G.~Pennycook, D.~Rothschild, {\it
  et~al.\/}, The science of fake news.
\newblock {\it Science\/} {\bf 359}, 1094--1096 (2018).

\bibitem{jacomy2014forceatlas2}
M.~Jacomy, T.~Venturini, S.~Heymann, M.~Bastian, Forceatlas2, a continuous
  graph layout algorithm for handy network visualization designed for the gephi
  software.
\newblock {\it PloS one\/} {\bf 9}, e98679 (2014).

\bibitem{blondel2008fast}
V.~D. Blondel, J.-L. Guillaume, R.~Lambiotte, E.~Lefebvre, Fast unfolding of
  communities in large networks.
\newblock {\it Journal of statistical mechanics: theory and experiment\/} {\bf
  2008}, P10008 (2008).

\end{thebibliography}
\end{document}



\baselineskip24pt


\maketitle

\section*{Introduction of an agent-based dynamical model without belief interaction mechanism between neighbors (Model 2)}

Here we present a model where nodes do not have their beliefs influenced by trusted discussion partners. Consider a network $G$ with $N$ nodes. Assume there are two competing viewpoints: viewpoint Blue ($\theta_1$) and viewpoint Red ($\theta_2$), where $\theta_1,\theta_2 \in [0,1]$. Each individual $i$ has its initial belief $w_i \in [0,1]$ which is uniformly distributed. At each timestep $T$, the following procedure occurs: 
\\[8pt]
$1$. An agent, $i$, is randomly selected.\\
$2$. Agent $i$ considers updating to either $\theta_1$ or $\theta_2$ with probability $\Omega_T$ and $1-\Omega_T$ respectively.\\
$3$. Assume $\theta_1$ is selected. One of four possible interactions occur:

    a. (Support candidate) If $\left|w_i-\theta_1\right|<\delta_1$, then $i$ becomes a Blue supporter. $i$ also becomes more ideologically similar to their new candidate, and $w_i$ is updated to $\bar{w_i}$ beliefs in accordance with a candidate influence parameter, $\mu_1$: $\bar{w_i}= (1-\mu_1) w_i+ \mu_1 \theta_1$, $\mu_1 \in [0,1]$.
    
    b. (Discuss with friend) If $\left|w_i-\theta_1\right|\ge\delta_1$ but $\left|w_i-w_j\right|<\delta_1$ and $\left|w_j-\theta_1\right|<\delta_1$, then $i$ is persuaded by $j$ to support the candidate and updates their beliefs in accordance with the candidate, like in $3a$.
    
    c. (Stay put) If $\left|w_i-\theta_1\right|\ge\delta_1$,  $\left|w_i-w_j\right|<\delta_1$ and $\left|w_j-\theta_1\right|\ge\delta_1$, then $i$ retains their opinion and does not rewire. 
        
    d. (Find new friend) If $\left|w_i-\theta_1\right|\ge\delta_1$ and $\left|w_i-w_j\right|\ge\delta_1$ for the chosen candidate and social connection, then $i$ rewires their tie with connection $j$ to another tie, $k$, where $\left|w_i-w_k\right|<\delta_1$.
\\
$4$. Repeat step $1$ to $3$ until all agents update once.
\\

To sum up, now each node has four possible interactions within one updating round: the node becomes $S$ state and receives positive feedback from the viewpoint it supports; the node is advised by a friend who supports the viewpoint and its belief becomes closer to that viewpoint; the node finds that it holds similar opinions with its friend and does not make any change; or the node chooses to rewire (see schematic in Fig. S5B). The evolution moves to the next time step $T + 1$ when all notes update once. 

The dynamical processes stop when the system becomes steady. As can be seen in Fig. S5, the main difference of this new model lies in the absence of belief interactions between neighbors, i.e., peer influence. Under this circumstance, the direction of opinion evolution is more deterministic: when updates with viewpoint $\theta_v$, $v\in\{1,2\}$, the users’ beliefs either converge to $\theta_v$ or remain unchanged. Peer influence is what makes the process of opinion evolution stochastic, as it depends on the relative belief position of neighbors in the network.

Overall, Model 2 has three general similarities with the original model. First, in both models there are three phases of echo chamber emergence as a function of identity scope: many clusters, two clusters, and one cluster (Fig. S6). Second, in both models candidates can optimize their support by optimally countering the ideological viewpoint of their opponent (Fig. S8).  Finally, too much or too little campaign effort can leave certain individuals undecided (Fig. S9).

There are, however, some important differences of note. First and most importantly, Model 2 is more prone to extremist clusters of individuals who fall outside the identity scope of either candidate (Fig. S6C and F). From this we can conclude that peer influence, which is missing from Model 2, inhibits the formation of  small extremist clusters (and possibly clusters nested between the two alternatives, if the identity scope is small enough). Second, levels of candidate influence ($\mu_1$) non-monotonically shapes proportions of supporters (Fig. S7). When candidate influence is low, individuals are largely unmoved by candidates, but when candidate influence is too strong it hinders the sufficient social discussion. Finally, we notice that the simulation results of Model 2 are much smoother in all figures, which is explained by the determinism of the model in the absence of peer influence.

In conclusion, comparing the results of Model 2 with findings from the main text shows that the stochasticity introduced by peer influence, while making things more unpredictable, also prevents the formation of small clusters of extremists. Both models are informative, and each may correspond to some situations and environments better than others. For example, in a society with lots of top-down political discussion from debates on networks and political advertisements, the positions of candidates may be more important and Model 2 may be appropriate. Whereas a culture with a culture of decentralized political discussion may be better reflected by the model in the main text. An exploration of the two-dimensional (top-down and peer-to-peer) space of political influence may be fertile ground for future research.

\clearpage
\newcommand{\beginsupplement}{%
        \setcounter{table}{0}
        \renewcommand{\thetable}{S\arabic{table}}%
        \setcounter{figure}{0}
        \renewcommand{\thefigure}{S\arabic{figure}}%
     }
     
\beginsupplement     


\begin{figure*}[h]
\centering
\includegraphics[width=\linewidth]{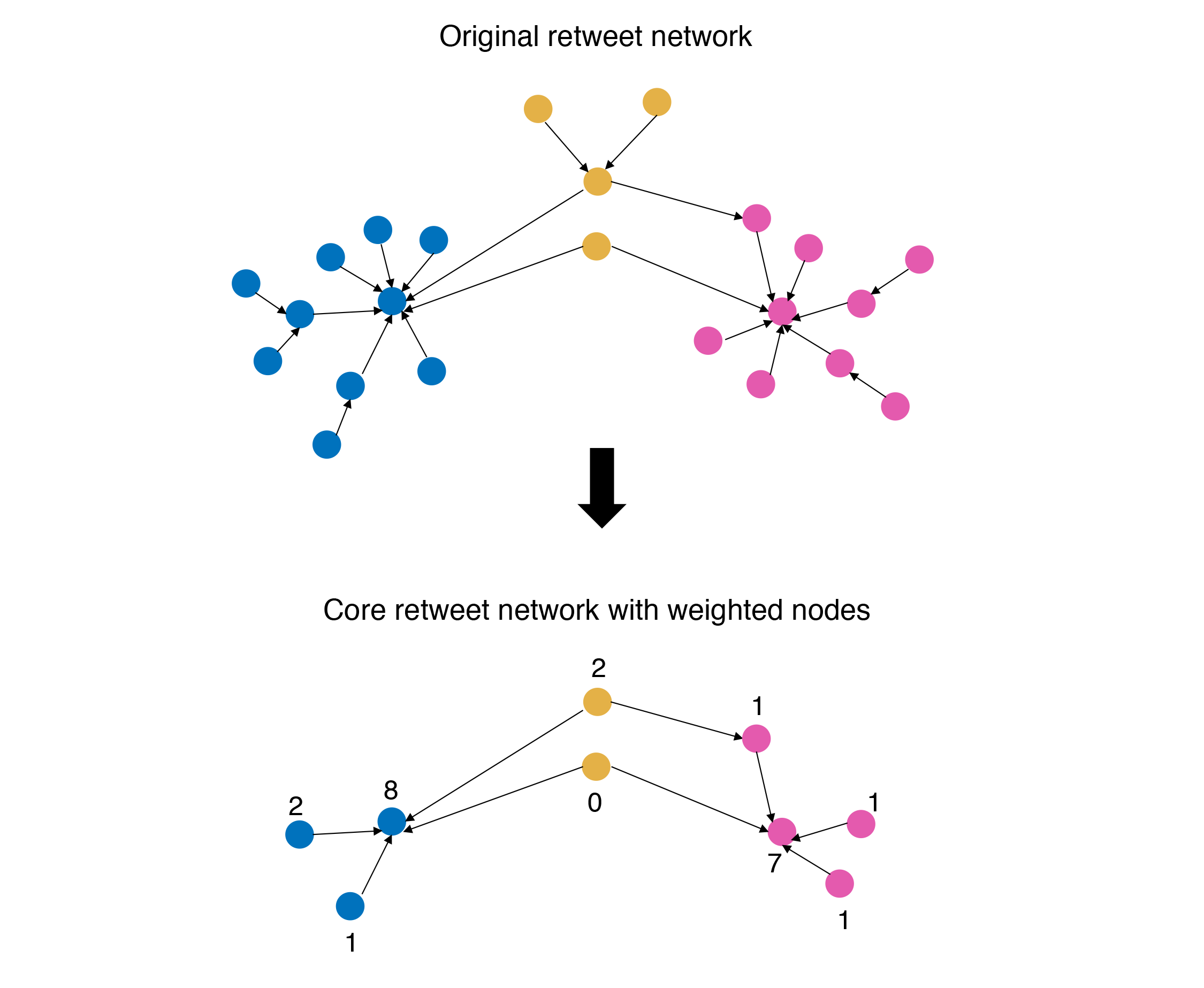}
\caption{ \textbf{Data processing method of retweet networks}. Define the weight of each node as the total number of times it is retweeted. In the largest connected component, we get rid of those ``pure followers" who are not being retweeted (in-degree equals to $0$) and only retweet others once (out-degree equals to $1$) and plot the remaining retweet network with all nodes' weights recorded. The simplified retweet network has all the important information and better represents the core structural characters.}
\label{figS0}
\end{figure*}
\clearpage

\begin{figure*}[ht]
\centering
\includegraphics[width=\linewidth]{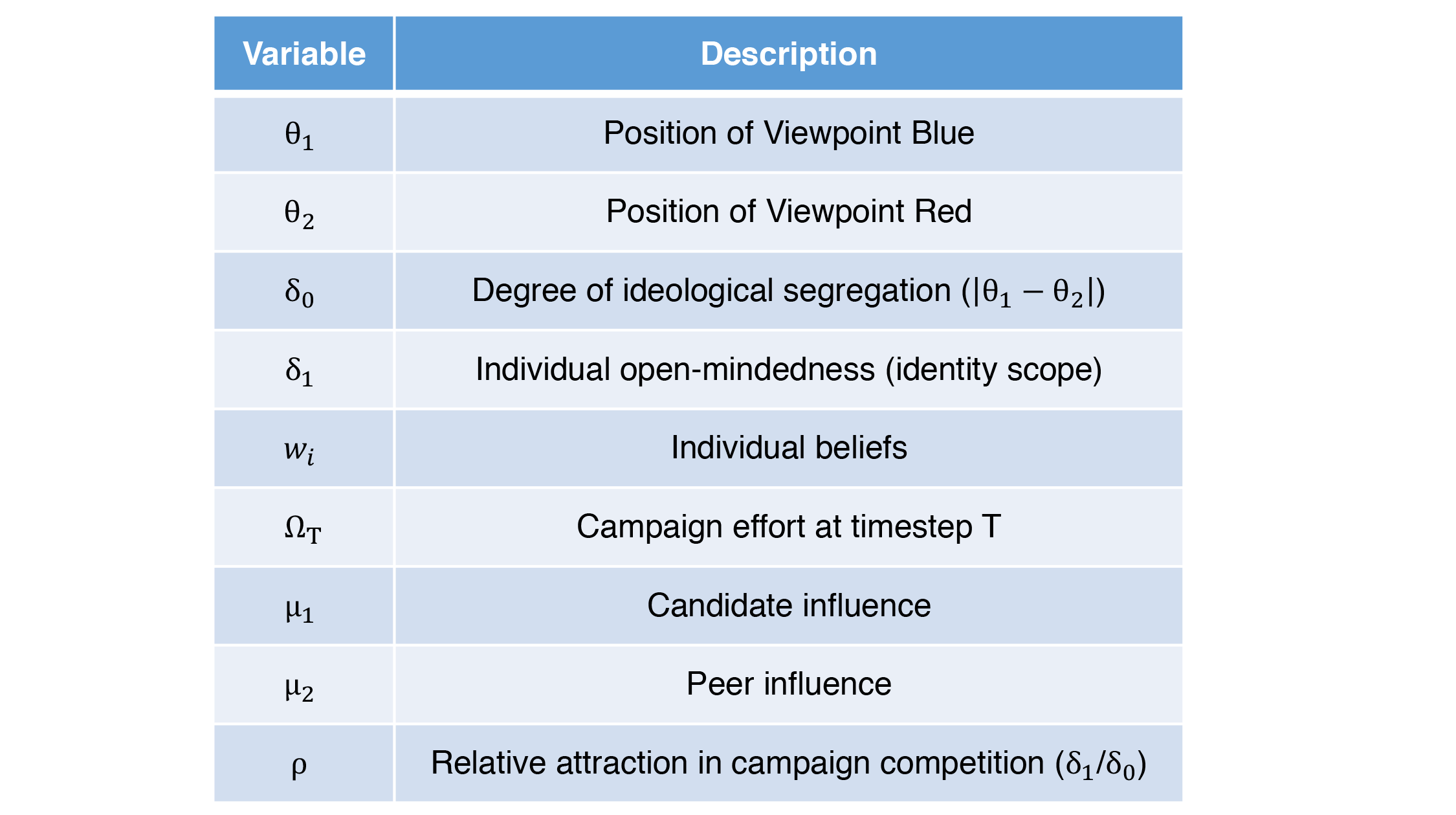}
\caption{ \textbf{List of model parameters}. Model variables and their descriptions.}
\label{figS2}
\end{figure*}
\clearpage

\begin{figure*}[ht]
\centering
\includegraphics[width=0.95\linewidth]{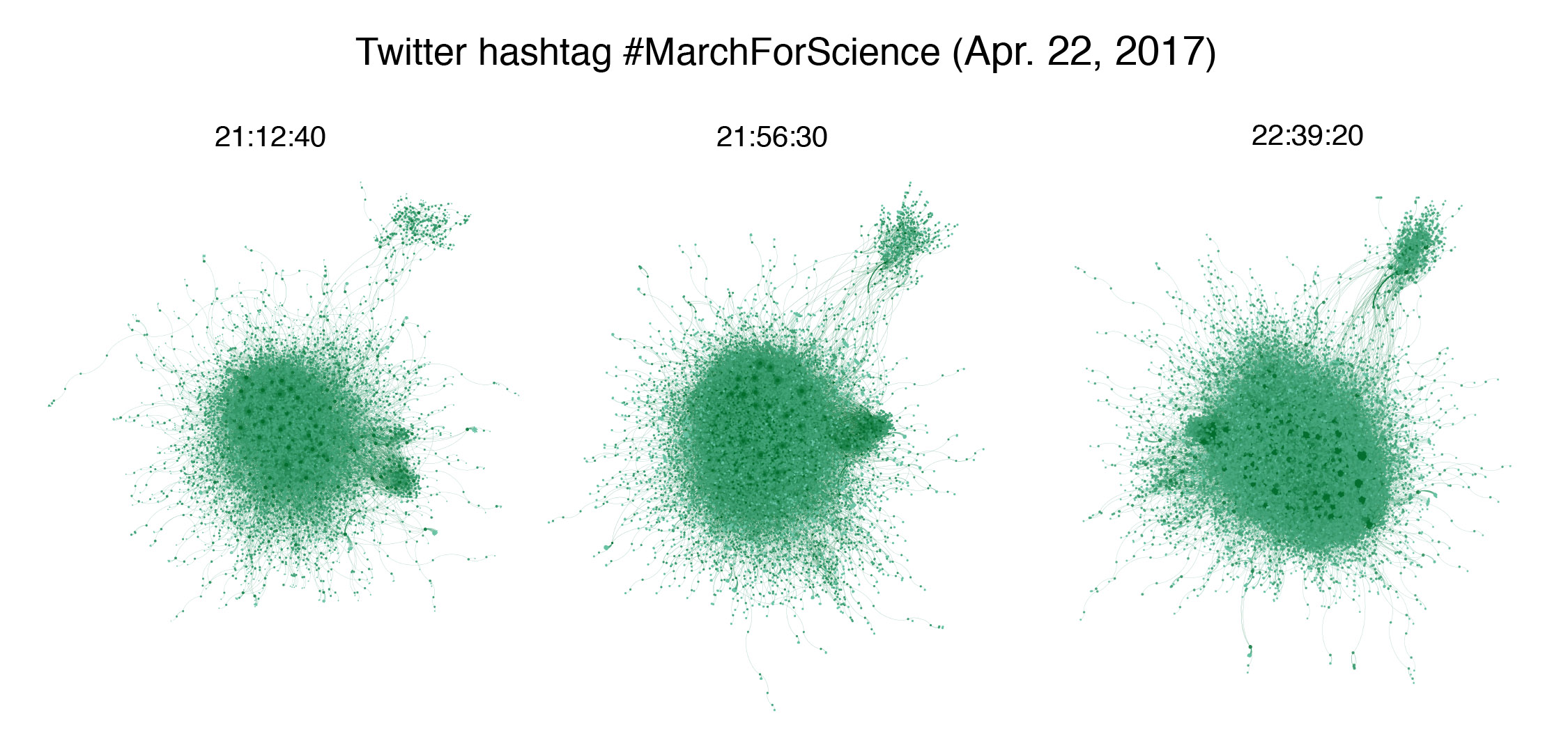}
\caption{ \textbf{Retweet network evolution of Twitter hashtag \#MarchForScience}. For comparison, we provide the existence of real-world event that has one giant cluster (as opposed to polarization).}
\label{figS1}
\end{figure*}
\clearpage

\begin{figure*}[ht]
\centering
\includegraphics[width=0.95\linewidth]{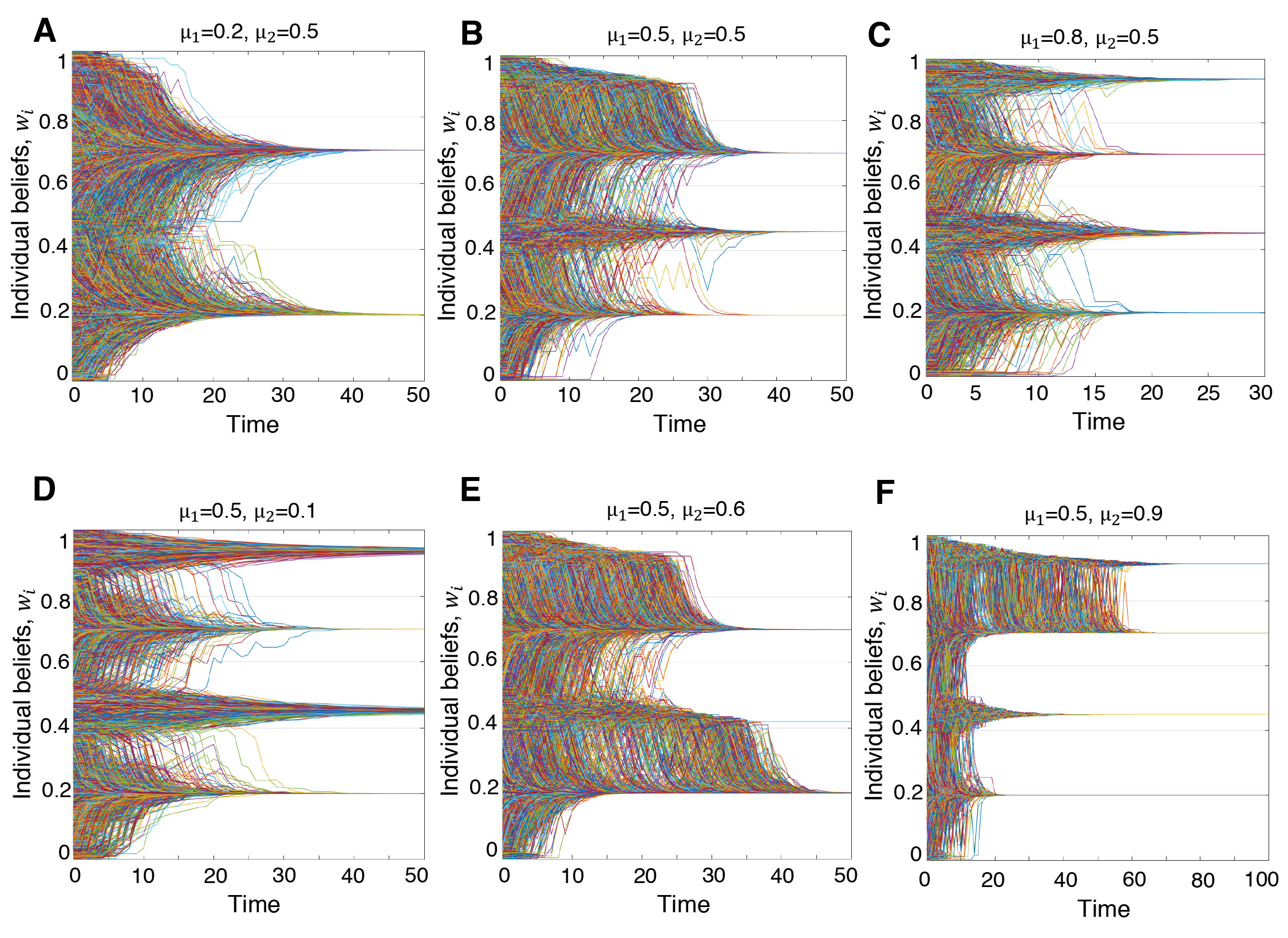}
\caption{ \textbf{Evolution of individual beliefs under different degrees of confirmation bias}. (A to C) Strong positive feedback between candidates and voters accelerates the opinion polarizing process and leads to insufficient social discourse where more undecided individuals exist. (D to F) Intermediate levels of discussion partner influence increase interactions between individuals of different beliefs, which reduces the undecided people and effectively prevents the formation of small clusters of extremists. Parameters: all simulations begin from an ER graph with $N=10^4$, $\langle k \rangle=6$ and $\Omega_T=0.5,$ $\theta_1=0.2,$ $\theta_2=0.7,$ $\delta_1=0.18$. }
\label{figS3}
\end{figure*}
\clearpage

\begin{figure*}[ht]
\centering
\includegraphics[width=0.95\linewidth]{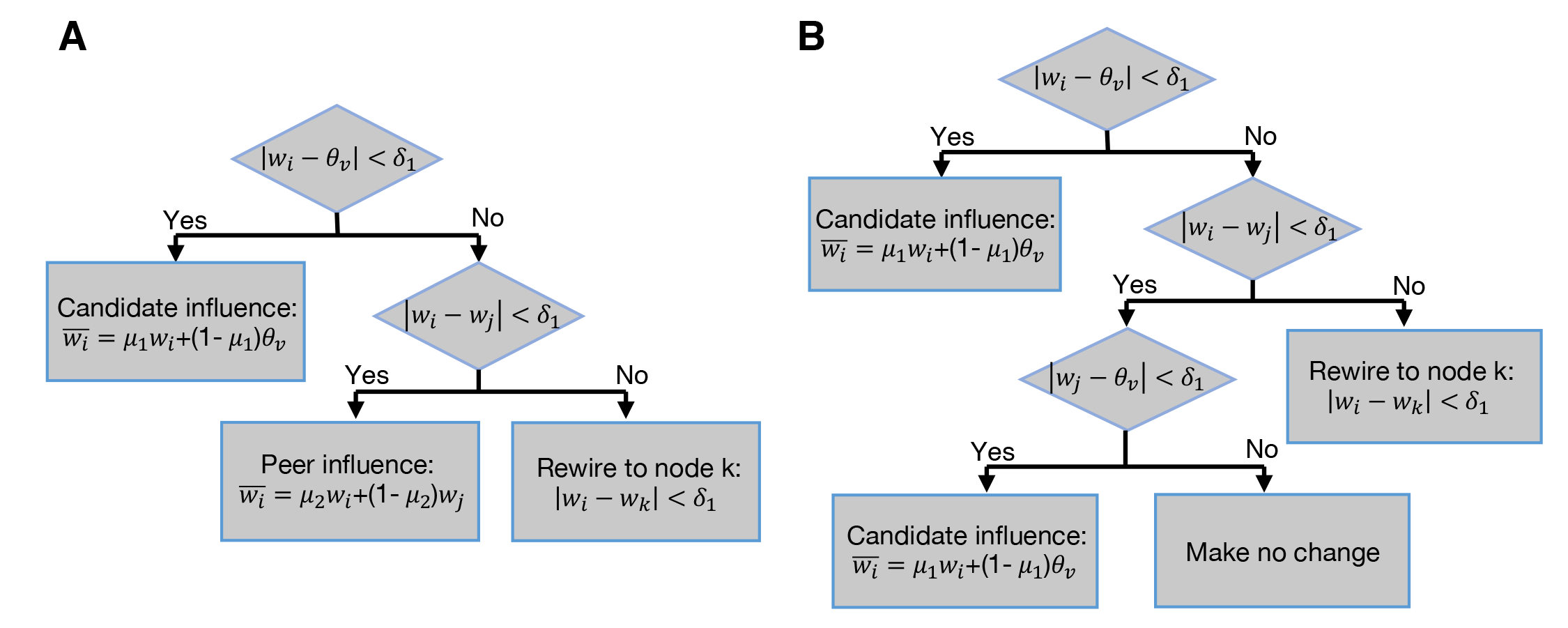}
\caption{ \textbf{Schematic of the two model frameworks}. (\textbf{A}) Framework of the dynamic agent-based model in the main text. (\textbf{B}) Framework of the new dynamical evolution model (Model 2) without discussion partner influence, i.e., there is no influence between neighboring voters.}
\label{figS4}
\end{figure*}
\clearpage

\begin{figure*}[htbp]
\centering
\includegraphics[width=0.95\linewidth]{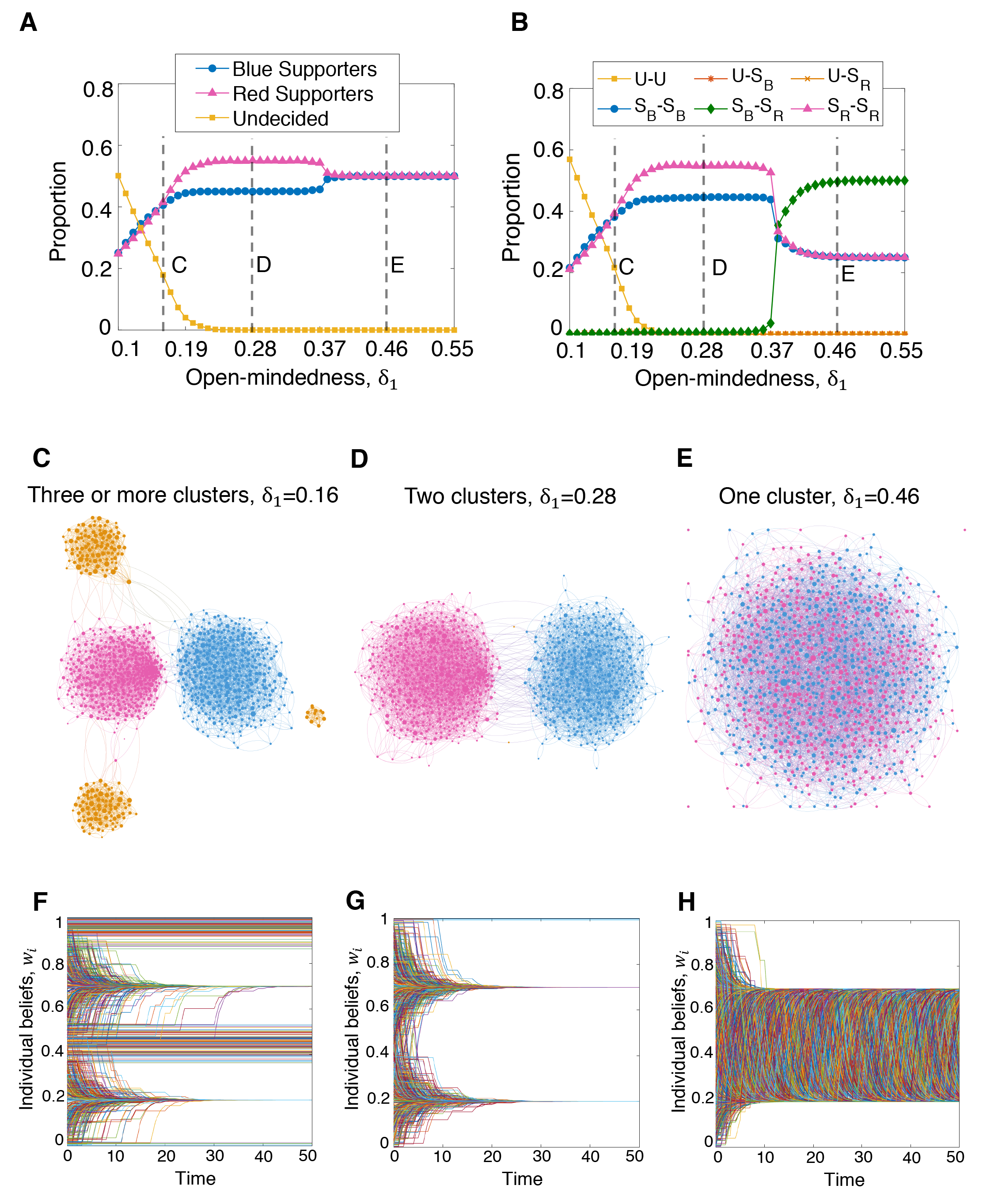}
\caption{ \textbf{Modeling the emergence of echo chambers and individual belief evolutions in Model 2.} The (\textbf{A}) breakdown of candidate support and (\textbf{B}) the distribution of ties between and within supporter groups indicate three possible opinion formation outcomes. The corresponding stable network structures and individual belief evolutions are present in (C to E) and (F to H), respectively. Of particular interest, the ``extremist" clusters are more likely to form and their beliefs will rarely change during the whole evolution processes (C and F). Parameters: $\Omega_T=0.5,$ $\theta_1=0.2,$ $\theta_2=0.7$, $\mu_1=0.5$. All evolutions begin from an ER graph with (A to B) $N=10^4$, $\langle k \rangle=6$. Simulation results are averaged over $100$ independent runs. In (C to H), $N=10^3$ and $\langle k \rangle=6$. }
\label{figD1}
\end{figure*}
\clearpage

\begin{figure*}[ht]
\centering
\includegraphics[width=0.95\linewidth]{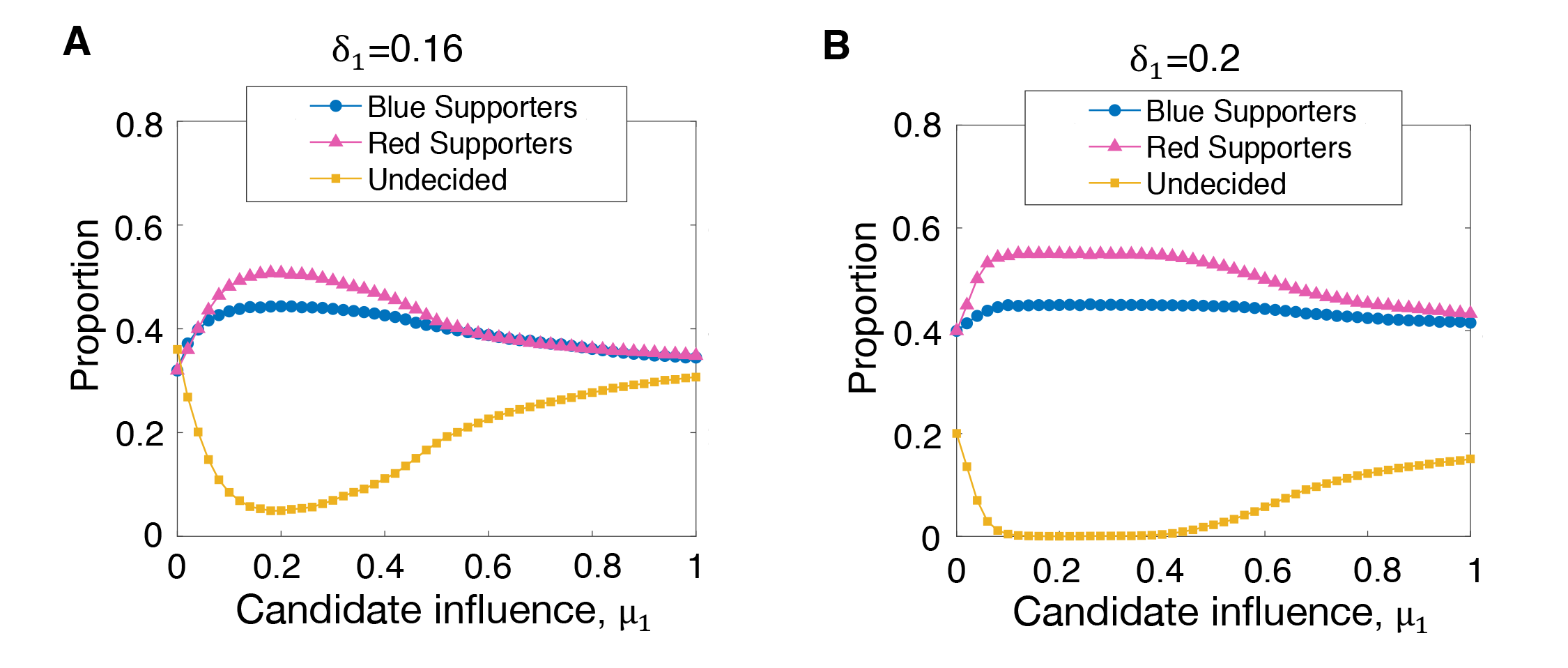}
\caption{ \textbf{Candidate influence and attitude evolution in Model 2.} We show how candidate influence without the influence of discussion partners, which is described by solely $\mu_1$, influences opinion formation. All simulation begins from an ER graph with $N=10^4$ with $\langle k \rangle=6$. Simulation results are averaged over $100$ independent runs. Parameters: $\Omega_T=0.5,$ $\theta_1=0.2,$ $\theta_2=0.7.$ (A) $\delta_1=0.16,$ (B) $\delta_1=0.2$.}
\label{figD2}
\end{figure*}
\clearpage

\begin{figure*}[ht]
\centering
\includegraphics[width=0.95\linewidth]{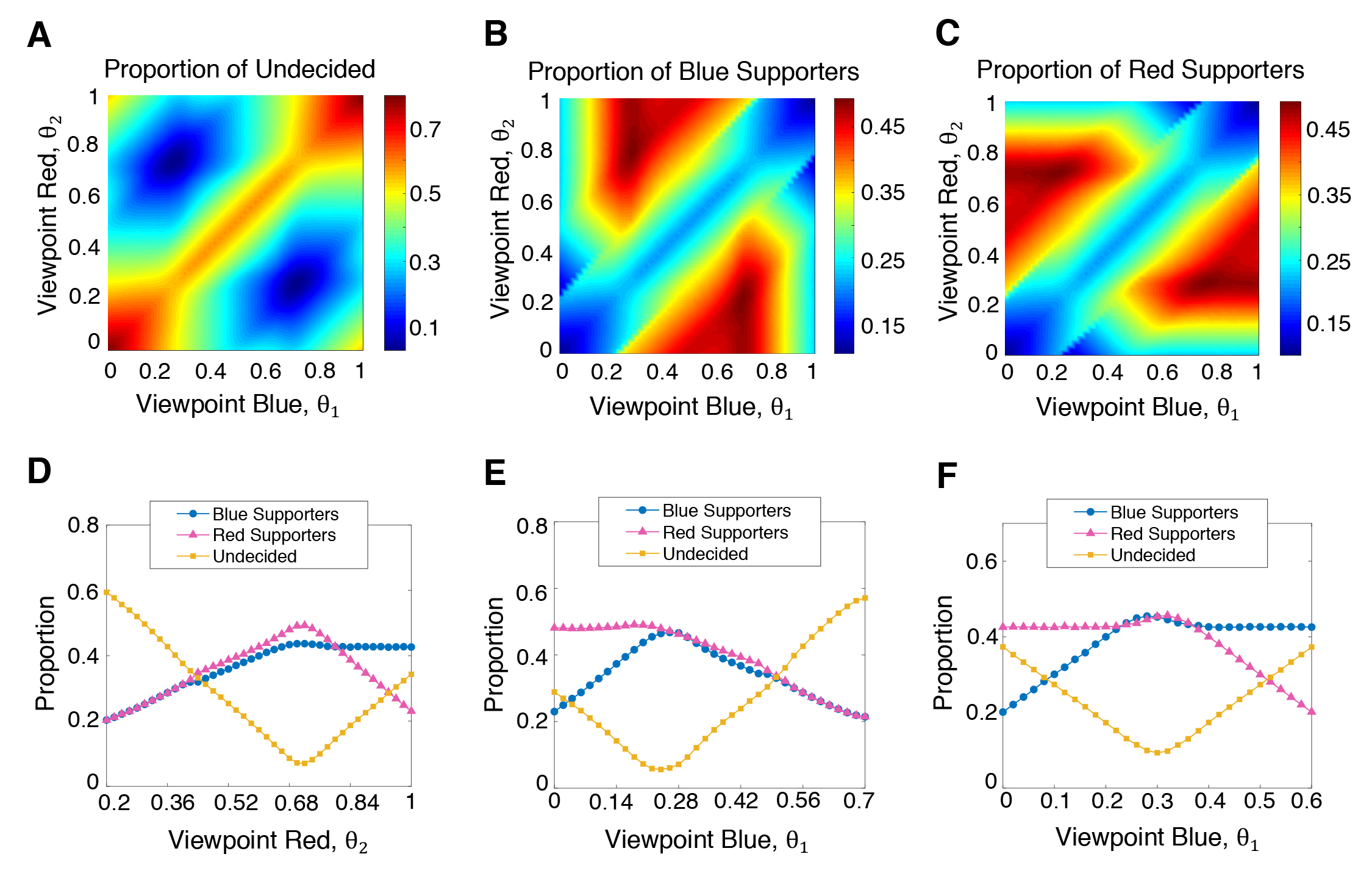}
\caption{ \textbf{``Attractiveness" of campaign viewpoint can as a function of candidate ideologies and socio-cognitive biases (Model 2).} All simulations begin from an ER graph with $N=10^4$ with $\langle k \rangle =6$. Simulation results are averaged over $100$ independent runs. Parameters: $\Omega_T=0.5,$ $\mu_1=0.5,$ $\delta_1=0.18$. (D) $\theta_1=0.2$; (E) $\theta_2=0.7$; (F) $\delta_0=0.4$. }
\label{figD3}
\end{figure*}
\clearpage

\begin{figure*}[ht]
\centering
\includegraphics[width=0.95\linewidth]{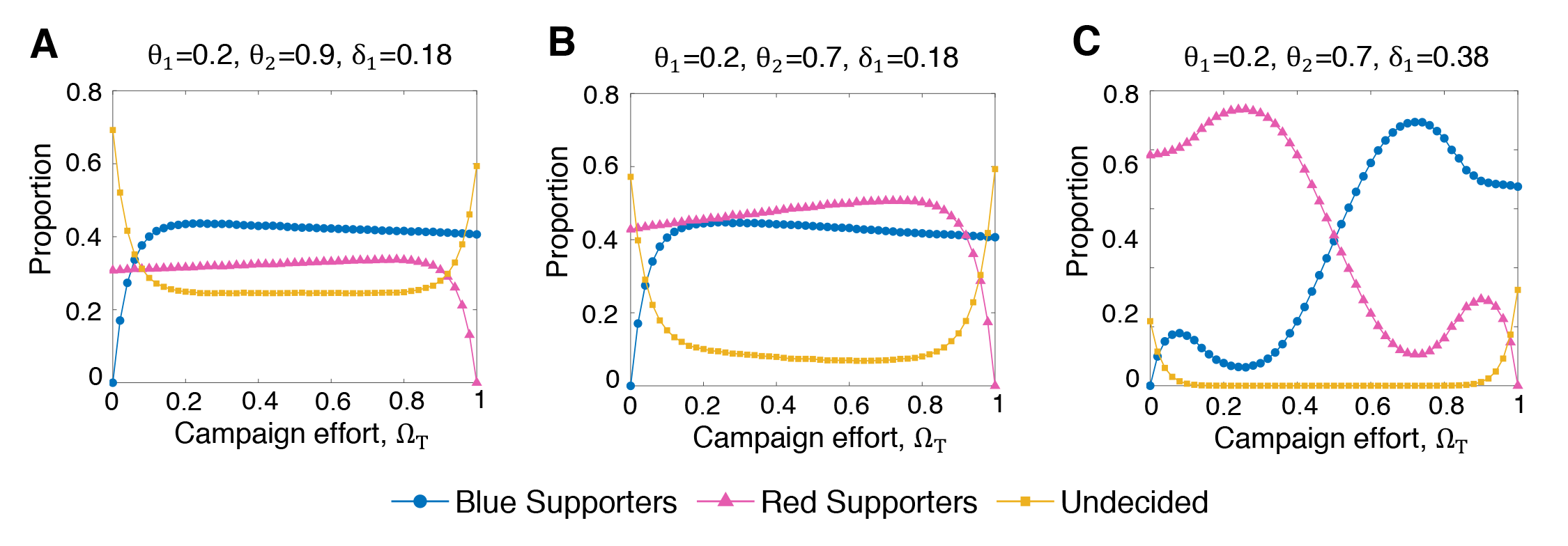}
\caption{ \textbf{Impact of campaign effort on winning majority support in Model 2.} Parameters: all simulations begin from an ER graph with $N=10^4$, $\langle k \rangle =6$ and $\mu_1=0.5$. Simulation results are averaged over $100$ independent runs. Relative attraction of campaign viewpoints: (A) $\rho=0.257$, (B) $\rho=0.36$, (C) $\rho=0.76$. }
\label{figD3}
\end{figure*}